\documentclass[journal]{IEEEtran}
\usepackage{scalerel}
\usepackage{tikz}
\usetikzlibrary{svg.path}
\definecolor{orcidlogocol}{HTML}{A6CE39}
\tikzset{
  orcidlogo/.pic={
    \fill[orcidlogocol] svg{M256,128c0,70.7-57.3,128-128,128C57.3,256,0,198.7,0,128C0,57.3,57.3,0,128,0C198.7,0,256,57.3,256,128z};
    \fill[white] svg{M86.3,186.2H70.9V79.1h15.4v48.4V186.2z}
                 svg{M108.9,79.1h41.6c39.6,0,57,28.3,57,53.6c0,27.5-21.5,53.6-56.8,53.6h-41.8V79.1z M124.3,172.4h24.5c34.9,0,42.9-26.5,42.9-39.7c0-21.5-13.7-39.7-43.7-39.7h-23.7V172.4z}
                 svg{M88.7,56.8c0,5.5-4.5,10.1-10.1,10.1c-5.6,0-10.1-4.6-10.1-10.1c0-5.6,4.5-10.1,10.1-10.1C84.2,46.7,88.7,51.3,88.7,56.8z};
  }
}
\newcommand\orcidicon[1]{\href{https://orcid.org/#1}{\mbox{\scalerel*{
\begin{tikzpicture}[yscale=-1,transform shape]
\pic{orcidlogo};
\end{tikzpicture}
}{|}}}}
\usepackage{hyperref}
\IEEEoverridecommandlockouts
\usepackage{blindtext}
\usepackage{lipsum}
\usepackage{multicol}
\usepackage{cite}
\usepackage{pgfplots}

\usepackage{adjustbox}
\usepackage{amsmath,amssymb,amsfonts}
\usepackage{algorithm}
\usepackage{algpseudocode}
\usepackage{subcaption}
\usepackage{graphicx}
\usepackage{textcomp}
\usepackage{enumitem}
\usepackage{xcolor}
\usepackage{mwe}
\usepackage{changepage}
\usepackage{caption}
\usepackage{subcaption}
\usepackage[utf8]{inputenc}
\usepackage{multirow}
\usepackage{xcolor}
\usepackage{framed}
\usepackage[normalem]{ulem}
\colorlet{shadecolor}{yellow!70}
\def\BibTeX{{\rm B\kern-.05em{\sc i\kern-.025em b}\kern-.08em
    T\kern-.1667em\lower.7ex\hbox{E}\kern-.125emX}}
\begin{document}

\title{ Reconfigurable Intelligent Surface-Assisted Cross-Layer Authentication for Secure and Efficient Vehicular Communications
}
\author{ Mahmoud~A.~Shawky\orcidicon{0000-0003-3393-8460}\,,
Syed~Tariq~Shah\orcidicon{0000-0003-4722-1786}\,,~\IEEEmembership{ Member,~IEEE}, Ahmed~G.~Abdellatif\orcidicon{0000-0002-3440-8448}\,, Muhammad~A.~Imran\orcidicon{0000-0003-4743-9136}\,,~\IEEEmembership{Fellow Member,~IEEE}, Qammer~H.~Abbasi\orcidicon{0000-0002-7097-9969}\,, Shuja~Ansari\orcidicon{0000-0003-2071-0264}\,,\\~\IEEEmembership{Senior Member,~IEEE}, and Ahmad~Taha\orcidicon{0000-0003-1246-8981}\,,~\IEEEmembership{Member,~IEEE}
\thanks{M.A. Shawky, M.A. Imran, Q.H. Abbasi, S. Ansari, and A. Taha \{ mahmoud.shawky, Muhammad.Imran, Qammer.Abbasi, Shuja.Ansari, Ahmad.Taha\}@glasgow.ac.uk are with the James Watt School of Engineering, University of Glasgow, UK.
A.G. Abdellatif \{ag.abdellatef@zu.edu.eg\} is with the Department of Communications and Electronics Engineering, Air Defense College, EMA, Cairo, Egypt.
S.T. Shah \{syed.shah@essex.ac.uk\} is with the School of Computer Science and Electronic Engineering, University of Essex, Colchester, UK.}
}
\maketitle

\begin{abstract}
Intelligent transportation systems increasingly depend on wireless communication for broadcasting traffic messages and facilitating real-time vehicular communication. In this context, message authentication is crucial for establishing secure and reliable communication. However, security solutions must consider the dynamic nature of vehicular communication links, which fluctuate between line-of-sight~(LoS) and non-line-of-sight~(NLoS) due to obstructions. This paper proposes a lightweight cross-layer authentication scheme that employs public-key infrastructure (PKI)-based authentication for initial legitimacy detection/handshaking while using key-based physical-layer re-authentication for message verification. This approach reduces signature generation and signaling overheads associated with each transmission, thereby enhancing network scalability. However, the receiver operating characteristic (ROC; $P_d$: detection vs. $P_{FA}$: false alarm probabilities) of the latter decreases with lower signal-to-noise ratio~(SNR). To address this, we investigate the use of reconfigurable intelligent surfaces~(RISs) to strengthen the SNR directed toward the designated vehicle in shadowed areas (i.e., NLoS scenarios), thereby improving the ROC. Theoretical analysis and practical implementation are conducted using a $1$-bit RIS consisting of $64 \times 64$ reflective meta-surfaces. Experimental results show a significant improvement in $P_d$, increasing from $0.82$ to $0.96$ at SNR = $-6$ dB for an orthogonal frequency-division multiplexing (OFDM) system with $128$ subcarriers. We also conducted informal and formal security analyses using Burrows-Abadi-Needham (BAN) logic to prove the scheme’s ability to resist passive and active attacks. Furthermore, the proposed scheme reduces computational and communication overheads by 43\% and 13\%, respectively, compared to traditional cryptographic methods, demonstrating its superiority for real-time, challenging communication scenarios.
\end{abstract}

\begin{IEEEkeywords}
AVISPA simulation, BAN-Logic analysis, Cross-layer authentication, Public key infrastructure, Reconfigurable intelligent surface, Random oracle modelling. %\vspace{-0.1cm}
\end{IEEEkeywords}

\section{Introduction}\label{S1}
Road traffic accidents cause 1.35 million fatalities annually, resulting in about 3,700 deaths per day, and it is expected to rank fifth among the causes of death by 2030 \cite{b8}. To address this issue, the World Health Organization has recognised the importance of developing intelligent transportation systems that enable real-time communication from vehicle-to-vehicle (V2V) and vehicle-to-infrastructure (V2I) \cite{b9}. Vehicular ad-hoc networks (VANETs) generally consist of three primary components: a trusted authority (TA), roadside units (RSUs), and wireless communication devices located on vehicles, also known as onboard units (OBUs) \cite{b10}. Each vehicle transmits a safety-related message containing information on location, speed, and heading at a transmission rate of $ 100-300 \hspace{0.1cm} msec$ \cite{b11}. This significantly enhances the performance of many traffic-related applications, including safety, mobility, and autonomy. Moreover, it reduces the carbon footprint and facilitates green transportation. However, the open-access nature of wireless communication makes it vulnerable to typical attacks, such as impersonation and modification. Hence, message authentication is crucial in preventing such attacks \cite{b12}.

Generally, there are three common types of authentication in VANETs: public key infrastructure (PKI)-based, identity (ID)-based, and group signature (GS)-based \cite{b10}. In PKI-based authentication, each vehicle has a pair of private and public keys \cite{b13}. The private key is used to generate digital signatures on messages. For verification, the public key is attached to the transmitted message as a digital certificate signed by the TA. In ID-based authentication, the vehicle's identifier, such as the vehicle identification number, is used as its public key, which can verify signatures generated by the vehicle's private key. This approach eliminates the need for a separate public key infrastructure, as the identifier serves as the public key \cite{b14}. In GS-based authentication, group members generate the signature ($\sigma$) on behalf of the group using their secret keys, while the recipient verifies $\sigma$ using the group's public key \cite{b15}. The signature is generated so that it cannot be traced back to the specific member who generated it, offering anonymity and privacy preservation. However, these methods require complex cryptographic operations, leading to high computation and communication costs for transmitting and verifying messages.

To overcome this limitation, physical (PHY)-layer authentication techniques have emerged as a promising solution to reduce the overheads associated with upper-layer cryptographic approaches \cite{b16}. These techniques employ the unique features of wireless channels, such as channel amplitude and phase responses \cite{b17}, and the hardware impairments, such as analogue front-end imperfections and carrier frequency offset \cite{b18}, to discriminate between terminals. Interested readers in this topic are referred to \cite{b33, b34, b35, b36}, where Chen et al. \cite{b33, b34} introduce a secure PHY-layer message authentication mechanism, regardless of the computation availability of adversaries. The work presented in reference \cite{b35} introduces an innovative message authentication scheme that integrates a secure channel coding mechanism. This mechanism leverages random coding techniques to effectively identify potential man-in-the-middle (MITM) attacks. Reference \cite{b36} presents a keyless authentication methodology characterised by a high authentication rate, thereby enhancing network scalability. However, PHY-layer methods cannot be alternative to crypto-based methods due to challenges in extracting distinguishing features or channel attributes in hardware impairments-based and feature-based approaches. Additionally, observing all communicating terminals' secret features within a limited coherence period and dealing with minor differences between hardware impairments remain significant challenges. For more details, refer to \cite{b1}.

In this context, integrating the PHY-layer into the upper-layer authentication approaches enhances the network's security and scalability, introducing what is referred to as ``cross-layer authentication'' \cite{b19}. In VANETs, the concept of cross-layer authentication is inspired by human interaction dynamics, wherein individuals are initially identified and subsequently remembered based on distinct physical traits like facial attributes, body structure, vocal characteristics, and other distinguishing qualities. Similarly, dependable and secure communication within vehicular networks can be established through an initial handshake procedure that integrates cryptographic authentication. This pivotal step assumes a critical role in verifying the authenticity of participating vehicles and capturing distinctive features of wireless devices, which can subsequently serve for ongoing re-authentication during forthcoming transmissions \cite{b20}. By combining PHY-layer features with upper-layer cryptographic techniques, this cross-layer authentication approach significantly reduces both computational and communication overheads. Unlike traditional methods that require generating and transmitting digital signatures with every message, an often costly process in terms of processing time and bandwidth, this methodology limits complex cryptographic operations to the initial handshake phase. Subsequent message verifications rely on lightweight PHY-layer re-authentication, which exploits unique wireless channel characteristics inherent to each vehicle. This reduces the frequency and complexity of cryptographic computations and decreases the amount of additional signaling data exchanged across the network \cite{b1}. As a result, the approach enhances network scalability and responsiveness, which are critical for real-time vehicular communication systems operating under stringent resource constraints. However, the performance of PHY-layer-based techniques in terms of detection and false alarm probabilities depends on the signal-to-noise ratio (SNR). The higher the SNR, the higher the detection probability, and vice versa. Considering the significant wireless channel variations and the instability of vehicular communication links caused by unpredictable obstructions, the re-authentication performance can be adversely affected, posing a challenge.

Reconfigurable intelligent surfaces (RISs) has emerged as a novel class of planar meta-material structures that can manipulate and reflect incident electromagnetic waves through dynamic surface property adjustments \cite{b21}. By controlling the electromagnetic waves' reflection and scattering, RISs can enhance wireless communication systems' SNR values and improve PHY-layer re-authentication performance. Following is a summary of this paper's contributions:
\begin{enumerate}
    \item This paper proposes a novel pseudo-identity-based PHY-layer re-authentication scheme that complements the initial PKI-based legitimacy detection. This approach significantly reduces the need for computationally expensive cryptographic signature generation and transmission for every message, thereby reducing communication and computation overheads without compromising the security and privacy requirements of VANETs.
    \item To overcome the challenges posed by dynamic and obstructed vehicular communication channels, we introduce the use of RIS to actively strengthen the signal quality in NLoS scenarios. We rigorously verify this enhancement via both theoretical analysis and real-world experimentation using a $1$-bit RIS with $64 \times 64$ reflective units, demonstrating a significant improvement in detection probability from $0.82$ to $0.96$ at an SNR of $-6$ dB.
    \item We conduct comprehensive security analyses, including informal discussion and formal verification using Burrows-Abadi-Needham (BAN) logic, proving the scheme’s resilience against passive and active attacks. Furthermore, performance evaluations quantitatively show that our scheme reduces computational and communication overheads by 43\% and 13\%, respectively, compared to traditional cryptographic methods, underscoring its suitability for real-time VANET applications, particularly in challenging channel conditions.
\end{enumerate}

   The structure of the remainder of this paper is as follows. Section \ref{S2} reviews relevant prior works. Section \ref{S3} presents the proposed scheme. Section \ref{S4} analyses the scheme's security and privacy. Section \ref{S5} evaluates the scheme's performance. Finally, in Section \ref{S6}, we provide concluding remarks.
\section{Related works}\label{S2}
This section reviews existing work on traditional authentication approaches and cross-layer authentication.
\subsection{Traditional cryptographic signatures-based authentication}
This part reviews current authentication designs that seek to alleviate the significant overheads inherent in traditional authentication methods in VANETs. Liu et al.~\cite{b23} propose a proxy vehicle-based authentication scheme to mitigate the computational overhead on RSUs. This scheme adopts an ID-based approach, leveraging the computational resources of proxy vehicles to verify signatures on behalf of RSUs. However, Asaar et al.~\cite{b22} demonstrated that the scheme in~\cite{b23} is vulnerable to impersonation attacks. Unfortunately, in scenarios where proxy vehicles are unavailable, all signatures must be verified directly by RSUs, leading to increased computational and communication overheads. In \cite{b25}, Jiang et al. propose using region trust authorities to deliver efficient vehicle authentication services while reducing the computational load of TA and RSUs. In \cite{b24}, Lim et al. propose a GS-based solution that addresses the overheads of the TA by introducing a hierarchical structure of RSUs comprising leader and member RSUs. The leader RSUs are empowered to generate group keys as group managers, thereby reducing the overheads on the TA. However, a compromised leader RSU can compromise the security and privacy of group members within the region. In \cite{b26}, Shao et al. propose a batch verification-based authentication scheme that enables recipients to verify multiple received signatures simultaneously. However, the high computation complexity of bilinear pairing (BP) operations poses a significant challenge. Moreover, this method is susceptible to failure in the presence of a single invalid signature, which can lead to time-consuming singular verification.

Mohammed et al. \cite{b120} introduce a fog computing-based pseudonym authentication scheme utilising elliptic curve cryptosystem (ECC) and general hash functions to enhance privacy and security in fifth-generation~(5G)-enabled vehicular networks. In \cite{b3}, Cui et al. developed an ECC-based content-sharing scheme tailored for 5G-enabled vehicular networks. The authors' approach enables vehicles with content downloading requests to efficiently filter their nearby vehicles to select competent and suitable proxy vehicles. These selected proxy vehicles are then requested to provide content services. Wang et al. \cite{b4}, Li et al. \cite{b5}, and Almazroi et al. \cite{b121} proposed conditional privacy-preserving authentication schemes to reduce authentication overheads and promote privacy. By adopting these schemes, vehicles are not required to store any certificates for authentication, and the TA is also relieved of the need to retrieve the real identity of malicious vehicles from certificates. While the previously mentioned methods aim to achieve a higher authentication rate, many suffer from limited network scalability, highlighting the need for more effective solutions to meet the growing demands of vehicular communication systems.
\subsection{An overview of cross-layer authentication}
Wen et al. \cite{b19} patented a cross-layer authentication method that uses PKI-based authentication for handshaking and generates a radio frequency fingerprint for re-authentication. In \cite{b27, b28, b29}, PKI-based authentication is integrated with the PHY-layer re-authentication using the feature tracking mechanism. This mechanism depends on the spatial and temporal correlation of the wireless channel responses within the coherence period $T_c$. Mughal et al. \cite{b20} propose an approach for incorporating the integrated circuits (ICs) physically unclonable function (PUF). Based on the IC's physical variation ($P$), the PUF method effectively generates an unpredictable response $R = P(C)$, where $C$ is the input challenge. However, the scalability of hardware imperfections-based techniques is limited, as the false alarm probabilities increase with the number of terminals. This is due to the slight dissimilarities in the hardware impairments extracted features from different devices. For feature tracking-based techniques, the recipient must extensively observe all the communicating terminals' secret features within $T_c$, constituting a significant challenge. In this context, it is crucial to consider some parameters when selecting the PHY-layer re-authentication method. These include channel variations, broadcasting rates, computation availability, and communication ranges. In \cite{b1}, we presented two re-authentication mechanisms: the PHY-layer signature-based identity authentication mechanism (PHY-SIAM) and the PHY-layer feature-tracking mechanism (PHY-FTM). PHY-SIAM is a keyed-based PHY-layer authentication mechanism where the message content is hashed and encapsulated into two orthogonal frequency-division multiplexing (OFDM) symbols. This forms the PHY-layer signature that can only be equalised at the intended receiver's side. PHY-FTM is a feature-tracking mechanism that utilises the high correlation between the channel estimates of subsequent OFDM symbols to verify message integrity. However, the previous study presented in \cite{b1} focused on the re-authentication performance for subsequent transmissions, assuming that the initial authentication was conducted during the first time slot. 

\subsection{An overview of RIS-assisted PHY-layer authentication}
Recent studies have increasingly explored the integration of RISs into PHY-layer authentication (PLA), aiming to enhance security and robustness against adversarial attacks. These works demonstrate how RISs can be leveraged not only to improve wireless communication performance but also to provide new mechanisms for identity verification and attack resilience. For instance, Zhang et al.~\cite{b701} propose a lightweight PLA framework for IoT devices in smart cities based on tag embedding and verification, demonstrating strong resistance to active attacks and highlighting the scalability of RIS-assisted security solutions. Crosara et al.~\cite{b702} analyse divergence-minimizing attacks against challenge-response PLA (CR-PLA) in RIS-enabled environments. By modeling the adversary’s optimal strategy through Kullback–Leibler divergence, they expose vulnerabilities that emerge under different channel knowledge and SNR conditions, revealing the limits of CR-PLA. Similarly, Tomasin et al.~\cite{b703} introduce an environment-based CR-PLA paradigm where the verifier manipulates the electromagnetic environment, including RIS-assisted channels, to authenticate devices without explicit challenges, an approach that illustrates both opportunities and emerging threats.

Other works have focused on channel sparsity and cascaded features. Bendaimi et al.~\cite{b704} exploit the double-structured sparsity of RIS-assisted MIMO channels to design robust channel-based PLA, achieving significant improvements in detection and false alarm rates compared to conventional methods. Zhang et al.~\cite{b705} extend the tag-based approach by embedding cover tags constructed from intrinsic channel features, random signals, and cryptographic keys, validating their scheme’s resilience against impersonation and tampering. He et al.~\cite{b706} propose a generalised PLA scheme for RIS-assisted IoT that jointly exploits direct and cascaded channel features. Their analytical and numerical analysis shows enhanced detection performance across a wide range of system parameters. Beyond PLA, RIS has also been investigated for its role in key generation. Shawky et al.~\cite{b707} propose an RIS-enabled secret key generation framework for secure vehicular communications under denial-of-service attacks, showing that RIS can simultaneously support authentication and robust key extraction in adversarial environments. Despite these advances, most existing studies remain simulation-based and rely on idealised assumptions, such as requiring network nodes to be separated by at least $\lambda/2$ to ensure feature uniqueness. However, this assumption can be undermined if an attacker is able to closely shadow a legitimate device and capture highly correlated channel features, thereby compromising the security strength of PLA. Furthermore, challenge-response mechanisms must typically operate within a short coherence interval $T_c$, which becomes particularly challenging in highly dynamic vehicular scenarios. To address these gaps, the present study proposes a comprehensive cross-layer authentication scheme that integrates PKI-based authentication for initial handshaking with a lightweight two-factor PHY-layer re-authentication process combining PHY-SIAM and PHY-FTM. Importantly, we also investigate how RIS can be utilised to further enhance re-authentication reliability, enhancing the detection probability of the PHY-layer re-authentication performance.

\section{RIS-assisted authentication: The proposed scheme}\label{S3}
This section describes the system model, discusses the proposed scheme in detail, and explains how the RIS enhances the scheme's performance at low SNR values.

\begin{figure}[t!]
\centerline{\includegraphics[width=8.8cm]{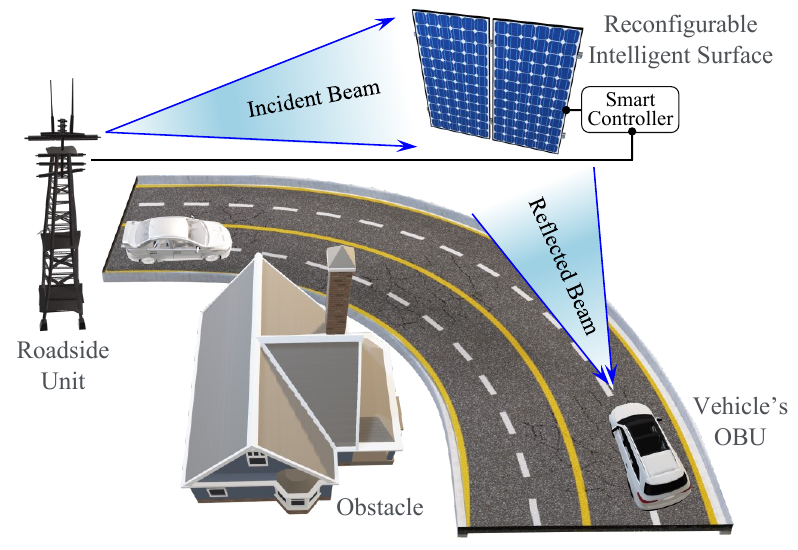}}
\setlength\belowcaptionskip{-0.5cm}
\caption{System modelling.}
\label{f1}
\end{figure}

\subsection{System modelling}
The system model of the proposed RIS-assisted vehicular communication scheme is depicted in Fig. \ref{f1}. The considered system model consists of the following entities.
\begin{itemize}
    \item \textit{TA}: The TA is a trusted entity for all network terminals, possessing sufficient computational resources to register and revoke any network terminal. It is also responsible for generating and distributing the system's public parameters. In addition, it is the only terminal capable of revealing vehicles' real identities in case of misbehaving (such as constructing an attack or violating traffic laws).
    \item \textit{RSU}: The RSU authenticates vehicles within range by verifying their broadcasted messages. It is also assumed to have a reliable communication link with the RIS's smart controller, where it can control the phase shift of the RIS elements. The RSU aims to optimise the RIS's configuration to form a directed beam toward the communicating vehicle in the shadowed areas. Additionally, it functions as a relay between communicating vehicles, extending the V2V communication range.
    \item \textit{Vehicle's OBU}: The OBU is a vehicle-mounted wireless communication device with limited computing capabilities. It can authenticate with nearby RSUs to send and receive real-time traffic conditions. We assume that both RSU and OBU are equipped with a single antenna.
    \item \textit{RIS}: The RIS comprises $L$ reconfigurable passive reflectors and is deployed to provide reliable communication links between the RSU and vehicles' OBUs (see Fig. \ref{f1}). By doing so, the reflected signal towards the designated vehicle/RSU can be deliberately strengthened or impaired. Each RIS has a smart controller that allows the RSU to adjust the phase shift of the RIS reflecting units by choosing between different configuration patterns.
\end{itemize}

The notations used in this paper are summarised in Table \ref{T2} for ease of understanding.
\begin{table*}[t!]
\caption{List of notations and their equivalent AVISPA symbols}
\begin{center}
\renewcommand{\arraystretch}{1} % Default value: 1
\begin{tabular}{l|l|l}
\hline \small\textbf{Symbol} &\small\textbf{Definition} &\small \textbf{AVISPA symbol} \\
\hline \small $PPs$ &  \small The system's public parameters &  \small -- \\
\hline \small $\beta$, $PK_{TA}$ &  \small The system's master key and TA's public key &  \small \_inv\{PKTA\}, PKTA \\
\hline \small $PK_{V_i}$, $SK_{V_i}$ &  \small $V_i$'s certificate public and private keys &  \small PKV1, \_inv\{PKV1\} \\
\hline \small $Cert_{V_i}$ &  \small $V_i$'s long-term digital certificate &  \small PKV1.TR.\{PKV1.TR\}\_inv(PKTA)\\
\hline \small $T_R$ &  \small The certificate validation time &  \small TR\\
\hline \small $SPK_{V_i}$, $SSK_{V_i}$ & \small $V_i$'s session public and private keys &  \small  SPKV1, \_inv\{SPKV1\} \\
\hline \small $TID_{V_i}$, $PID_{V_i}$ &  \small $V_i$'s temporary and pseudo identities &  \small TIDV1, PIDV1 \\
\hline \small $PK_{R_k}$, $SK_{R_k}$ &  \small $R_k$'s  certificate public and private keys &  \small PKRSU, \_inv\{PKRSU\}\\
\hline \small $Cert_{R_k}$ &  \small $R_k$'s long-term digital certificate &  \small PKRSU.TR.\{PKRSU.TR\}\_inv(PKTA)\\
\hline \small $SPK_{R_k}$, $SSK_{R_k}$ &  \small $R_k$'s session public and private keys &  \small SPKRSU, \_inv\{SPKRSU\}\\
\hline \small $TID_{R_k}$ &  \small $R_k$'s temporary identities &  \small TIDRK\\
\hline \small $Sk_{i-k}$ &  \small The shared key between $V_i$ and $R_k$ &  \small SK12\\
\hline \small $\sigma_{V_i}$, $\sigma_{R_k}$ &  \small $V_i$'s and $R_k$'s signatures &  \small \{--\}\_inv\{PKV1\}, \{--\}\_inv\{PKRSU\}\\
\hline \small $\sigma_{V_i}^{PHY}$ &  \small $V_i$'s PHY-layer signature &  \small \{--\}\_SK12\\
\hline \small $\phi_a$, $\phi_b$ &  \small The PHY-layer signature's phase shifts &  \small --\\
\hline \small $T_i$, $T_r$ &  \small signatures' creating and receiving timestamps &  \small Ti $\forall$i $=\{1, 2, 3\}$\\
\hline \small $T_\Delta$ &  \small Timestamps' expiration period, e.g., [00:00:59] &  \small --\\
\hline \small $P_d$, $P_{fa}$ &  \small The detection and false alarm probabilities &  \small --\\
\hline
\end{tabular}
\vspace{-0.5cm}\label{T2}
\end{center}
\end{table*}

\subsection{The proposed authentication scheme}
This section provides a detailed discussion of the proposed scheme. In this work, each terminal has a long-term digital certificate for initial verification and handshaking between two legitimate parties. For re-authentication and secure message verification between vehicles and RSUs, we use PHY-SIAM and PHY-FTM \cite{b1} as a two-factor re-authentication method for the OFDM system of $N$ subcarriers. The handshaking process draws inspiration from conventional PKI-based mutual authentication, it significantly extends its capabilities through RIS-aware integration while satisfying the security and privacy requirement. These enhancements enable lightweight, low-latency, and physically resilient authentication tailored for dynamic vehicular environments. Generally, the proposed scheme comprises four phases, i.e., initialisation, registration, initial authentication, and message signing and verification.

\begin{table}[!t]
\caption{The $160$-bit $EC$'s recommended parameters of “$secp160k1$” in the Hexadecimal form \cite{b2}}
\begin{center}
\begin{tabular}{l|l}
\hline \footnotesize	\textbf{Par.} & \footnotesize	\textbf{Value}\\
\hline  \small $a$ &  \scriptsize	 $00000000\hspace{0.1cm}00000000\hspace{0.1cm}00000000\hspace{0.1cm}00000000\hspace{0.1cm}00000000$ \\
\hline  \small	$b$ &  \scriptsize	$00000000\hspace{0.1cm}00000000\hspace{0.1cm}00000000\hspace{0.1cm}00000000\hspace{0.1cm}00000007$ \\
\hline  \small	$p$ &  \tiny	$FFFFFFFF\hspace{0.1cm}FFFFFFFF\hspace{0.1cm}FFFFFFFF\hspace{0.1cm}FFFFFFFE\hspace{0.1cm}FFFFAC73$ \\
\hline  \small	$q$ & \scriptsize	 $01\hspace{0.1cm}00000000\hspace{0.1cm}00000000\hspace{0.1cm}0001B8FA\hspace{0.1cm}16DFAB9A\hspace{0.1cm}CA16B6B3$\\
\hline  \small	$P$ &  \scriptsize $04\hspace{0.1cm}3B4C382C\hspace{0.1cm}E37AA192\hspace{0.1cm}A4019E76\hspace{0.1cm}3036F4F5\hspace{0.1cm}DD4D7EBB$\\
 \small	 &  \scriptsize	 $938CF935\hspace{0.1cm}318FDCED\hspace{0.1cm}6BC28286\hspace{0.1cm}531733C3\hspace{0.1cm}F03C4FEE$\\
\hline
\end{tabular}
\vspace{-0.5cm}
\label{T1}
\end{center}
\end{table}
\subsubsection{System initialisation phase}
The TA follows the following steps to initialise the system's public parameters.
\begin{itemize}
    \item The scheme is designed based on the $80$-bit security level of the elliptic curve $E: y^{2}=x^{3}+a x+b \bmod p$. In this context, we adopted the $160$-bit elliptic curve, which is parameterised using the recommended domain settings of ``$secp160k1$'' \cite{b2}, as listed in Table \ref{T1}.
    \item Based on the generator $P$, the TA generates a cyclic group $\mathbb{G}$ of order $q$, which consists of all $E$'s points as well as the infinity point $\mathcal{O}$.
    \item The TA chooses the system master key $\beta \in Z_q^*$, then computes its related public parameter $PK_{TA}=\beta.P$.
    \item The TA selects two hash functions $H_1:\{0,1\}^*\rightarrow\{0,1\}^{N_{1}}$ and $H_2:\{0,1\}^*\rightarrow\{0,1\}^{2N_2}$ for $N_2=3N/4$. It also selects the $2$-bit Gray code mapping function $\mathcal{M}(x_{i})\rightarrow\phi_i$ that maps $x_{i}$ to $\phi_i$ as follows.
    \begin{equation}\label{e1}
    \phi_{i}=\mathcal{M}\left(x_{i}\right)= \begin{cases}0 & x_{i}=[00] \\ \frac{\pi}{2} & x_{i}=[01] \\ \pi & x_{i}=[11] \\ \frac{3 \pi}{2} & x_{i}=[10]\end{cases}, \forall i \in[1, N_2]
    \end{equation}
    \item Finally, the system's public parameters $P P s$ can be represented by the tuple $\langle a, b, p, q, P, Pk_{TA},$ $ H_1, H_2, \mathcal{M}\rangle$.
    \end{itemize}
\subsubsection{Registration phase}
    The TA registers all terminals before being part of the network by performing the following steps.
    \begin{itemize}
        \item For vehicle registration, the TA checks the vehicle $V_i$'s real identity $RID_{V_i}$, selects at random $V_i$'s secret key $SK_{V_i} \in Z_q^*$, and calculates its related public parameter $PK_{V_i}=SK_{V_i}.P$. Finally, the TA preloads the tuple $\langle P P s, SK_{V_i}, Cert_{V_i} \rangle$ onto $V_{i}$'s OBU, where $V_i$'s long-term digital certificate $Cert_{V_i}= \langle PK_{V_i}, T_R, \sigma_{TA} \rangle$, $\sigma_{TA}=Sign_{\beta}(PK_{V_i} \| T_R)$ and $T_R$ is the certificate validation time.
        \item Each RSU $R_k$ undergoes the same registration process.
        \item The TA creates a list of revoked vehicles' and RSUs' digital certificates known as the certificate revocation list $CRL=\{ Cert_{1}, ..., Cert_{z} \}$, where $z$ is the total number of revoked vehicles and RSUs. At last, the TA distributes the $CRL$ among vehicles via RSUs in different regions. The proposed scheme uses the traditional revocation mechanism of PKI-based authentication. However, the proposed method is designed to be compatible with contemporary contributions in revocation mechanisms, as outlined in \cite{b38}.
    \end{itemize}

\subsubsection{Initial authentication phase}
Consider a scenario where $V_i$ is within the communication range of $R_k$ and wants to initiate a secure connection. In this case, both terminals, $V_i$ and $R_k$, exchange certificate-based initial authentication packets for mutual legitimacy detection and extracting a symmetric shared key $SK_{i-k}$. The following steps constitute this phase.
\begin{itemize}
    \item $V_i$ randomly selects the session secret key $SSK_{V_i} \in Z_q^*$ and computes its corresponding public parameter $SPK_{V_i} = SSK_{V_i} . P$.
    \item $V_i$ selects at random a temporary identity $TID_{V_i} \in \{0,1\}^{N_{1}}$ and sends $R_k$ a request to communicate in the form of $\langle TID_{V_i}, SPK_{V_i}, T_1, Cert_{V_i}, \sigma_{V_i} \rangle$, where the signature $\sigma_{V_i}=\text{Sign}_{SK_{V_i}}(TID_{V_i} $ $\| SPK_{V_i} \| T_1 \| Cert_{V_i})$ and $T_1$ is the attached timestamp.
    \item Avoiding replay attacks, $R_k$ checks $T_1$'s freshness by testing whether if $T_r - T_1 \leq T_\Delta$ holds or not. Then, $R_k$ checks $V_i$'s legitimacy by determining if $Cert_{V_i} \in CRL$ holds or not. After that, $R_k$ authenticates the received tuple by verifying $\sigma_{V_i}$ as $\text{Verify}\hspace{0.1cm} (\sigma_{V_i})_{PK_{V_i}}$.
     \item In response to $V_i$'s request, $R_k$ selects at random $SSK_{R_k} \in Z_q^*$ and computes its corresponding public parameter $SPK_{R_k} = SSK_{R_k} . P$, computes $SK_{i-k}=SPK_{V_i}.SSK_{R_k}$ using the Diffie-Hellman key exchanging protocol, and sends the tuple $\langle TID_{R_k}, SPK_{R_k}, T_2, Cert_{R_k}, $ $\sigma_{R_k} \rangle$ to $V_i$, where $TID_{R_k}$ is the $R_k$'s temporary identity and $\sigma_{R_k}=\text{Sign}_{SK_{R_k}}(TID_{R_k} \|$ $ SPK_{R_k} \| T_2 \| Cert_{R_k})$.
    \item At last, $V_i$ checks if $T_r - T_2 \leq T_\Delta$ and $Cert_{R_k} \in CRL$ hold or not, verifies $\sigma_{R_k}$ as $\text{Verify}\hspace{0.1cm} (\sigma_{R_k})_{PK_{R_k}}$, and computes its own symmetric key $SK_{i-k}=SSK_{V_i}.SPK_{R_k}$.
    \item Each $R_k$ in a coverage area stores a list of communicating vehicles' temporary identities and their associated shared key so that $list_{R_k}=\{Tuple_1, ..., Tuple_n\}$, where $Tuple_i=\langle 
Cert_{V_i}, TID_{V_i}, SK_{i-k} \rangle$ $\forall i \in [1, n]$.
\end{itemize}
\begin{figure}[t!]
\centerline{\includegraphics[width=9cm]{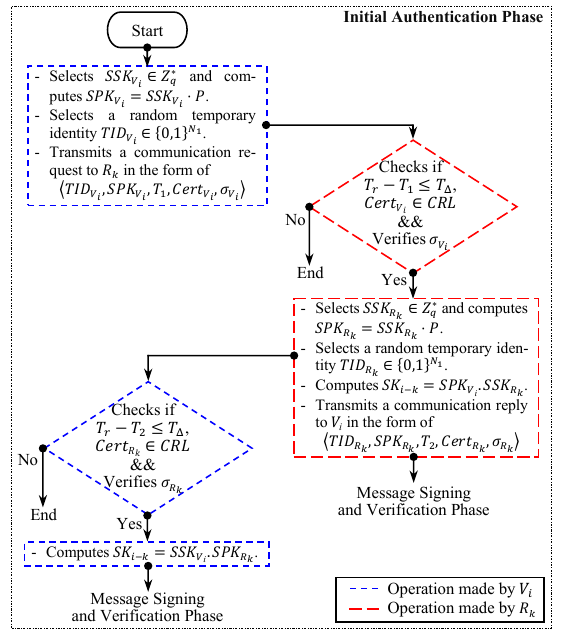}}
\setlength\belowcaptionskip{-0.5cm}
\caption{{The top-level description of the initial authentication.}}
\label{f2}
\end{figure}

Fig. \ref{f2} shows the top-level description flowchart of the initial authentication phase. Note that the same procedures can be reapplied for communication between two involved vehicles, $V_i$ and $V_k$, thereby facilitating V2V communication.

\subsubsection{Message signing and verification phase}
In this phase, we adopt PHY-SIAM and PHY-FTM proposed in \cite{b1} as a two-factor re-authentication process performed at the physical layer. We create a PHY-layer signature used as a lightweight re-authentication technique based on the symmetric shared key $SK_{i-k}$ and the message payload. Throughout this part, ${\mathbb{C}}^{N_x \times N_y}$, $\odot$, ${(\hspace{0.05cm})}^*$, and ${[\hspace{0.05cm}]}^T$ refer to a $N_x \times N_y$ matrix of complex elements, element-wise multiplication, conjugate, and transpose, respectively. While variables in uppercase and Bold represent matrices. The following steps constitute this phase.

\begin{itemize}
    \item For each specific number $Q$ of message ($m$) transmissions, $V_i$ selects a random $a_1 \in Z_q^*$ and calculates its related public parameter $A_1=a_1 . P$. Next, $V_i$ computes its pseudo-identity $P I D_{V_i}=T I D_{V_i} \oplus H_1(a_1 . P K_{R_k})$. 
    \item Then, $V_i$ sends $R_k$ the message in the form of $\langle m,$ $P I D_{V_i}, A_1, T_3, \sigma_{V_i}^{P H Y} \rangle$, where $\sigma_{V_i}^{P H Y}$ is the PHY-layer signature computed in a 2-step process as follows.\vspace{0.2cm}\\ 
    - \textit{Signature preparation step}: $V_i$ computes two OFDM symbols' phase shifts, $\mathbf{\Phi}_a=[e^{j{\phi_{a,1}}}, ...,$ $ e^{j{\phi_{a,N_2}}}]^T \in \mathbb{C}^{N_2 \times 1}$ and $\mathbf{\Phi}_b=[e^{j{\phi_{b,1}}}, ..., e^{j{\phi_{b,N_2}}}]^T \in \mathbb{C}^{N_2 \times 1}$, where $\phi_a=\mathcal{M}(H_2(\{SK_{i-k}\}_x \| T_3 \|$ $ A_1 \| P I D_{V_i} \| m ))$, $\phi_b=\mathcal{M}(H_2(\{SK_{i-k}\}_y \| $ $ T_3 \| A_1 \| P I D_{V_i} \| m ))$, and $\{.\}_x$ and $\{.\}_y$ represent the $x$ and $y$ coordinates of the elliptic curve point $SK_{i-k} \in \mathbb{G}$, respectively.\vspace{0.1cm}\\
    - \textit{Signature generation step}: In this step, $V_i$ encapsulates $\phi_a$ and $\phi_b$ onto two subsequent OFDM symbols of $N$ subcarriers and sends it to $R_k$ at times $t$ and $t+\Delta t$ so that the transmitted symbols can be represented as
    
\begin{equation}\label{e2}
\begin{aligned}
& \hspace{0.6cm}\mathbf{S}_{1}=[s_{1,1}, ..., s_{1,N_2}]^T=\mathbf{\Phi}_a \odot \mathbf{X}, \\
& \hspace{0.65cm}\mathbf{S}_{2}=[s_{2,1}, ..., s_{2,N_2}]^T=\mathbf{\Phi}_b \odot \mathbf{X}
\end{aligned}
\end{equation}
where $\mathbf{X} =[e^{j{\psi_1}}, ..., e^{j{\psi_{N_2}}}]^T \in \mathbb{C}^{N_2 \times 1}$, $\psi_i$ is a uniformly distributed random variable $\psi_i \sim U[0,2 \pi)$, and $\Delta t$ is the transmission time interval. Note that the OFDM symbols in (\ref{e2}) are collectively referred to as $\sigma_{V_i}^{P H Y}$. Also, we consider the OFDM system as a superposition of $N$ independently operating narrow-band subsystems.
\item $R_k$ receives $\sigma_{V_i}^{P H Y}$ in (\ref{e2}) at times $t^{\prime}$ and $t^{\prime}+\Delta t$, which can be represented in the frequency-domain, following the removal of the cyclic-prefix and calculating the Fast Fourier Transform (FFT), as
\begin{equation}\label{e3}
\begin{aligned}
& \mathbf{R}_{1}=[r_{1,1}, ..., r_{1,N_2}]^T=\left(\mathbf{H}_{VR} \odot \mathbf{S}_{1}\right)+\mathbf{N}, \\
& \mathbf{R}_{2}=[r_{2,1}, ..., r_{2,N_2}]^T=\left(\mathbf{H}_{VR}^\prime \odot \mathbf{S}_{2}\right)+\mathbf{N}^\prime
\end{aligned}
\end{equation}
    where $\mathbf{H}_{VR}=[|h_{1,1}|e^{j \xi_{1,1}}, ..., |h_{1,N_2}|e^{j \xi_{1,N_2}}]^T \in \mathbb{C}^{N_2 \times 1}$, $\mathbf{H}_{VR}^{\prime}=[|h_{1,1}^{\prime}|e^{j \xi_{1,1}^{\prime}}, ..., |h_{1,N_2}^{\prime}|e^{j \xi_{1,N_2}^{\prime}}]^T $ $\in \mathbb{C}^{N_2 \times 1}$, $\{|h_{1,i}|, \xi_{1,i}\}$ and $\{|h_{1,i}^{\prime}|, \xi_{1,i}^{\prime}\}$ are the channel amplitude and phase responses of the $i^{th}$ subcarrier at times $t^{\prime}$ and $t^{\prime}+\Delta t$, respectively, and $\{\mathbf{N}, \mathbf{N}^{\prime}\}$ are complex additive Gaussian noises $\mathbb{C} \mathbb{N}(0, \sigma_n^2)^{N_2 \times 1}$ with means and variances equal zero and $\sigma_n^2$, respectively. Note that $\mathbf{H}_{VR}$ is highly correlated with $\mathbf{H}_{VR}^\prime$ for $\Delta t \leq T_c$.
    \item $R_k$ checks $T_3$'s freshness, computes $TID_{V_{i}}= PID_{V_{i}}$ $ \oplus H_1(A_1 . S K_{R_k})$ and finds out if $TID_{V_{i}} \in list_{R_k}$ holds or no. If yes, $R_k$ uses $SK_{i-k}$ associated with $TID_{V_{i}}$ and the message payload $\langle m, P I D_{V_i}, A_1, T_3 \rangle$ to compute $\phi_a^{\prime}=\mathcal{M}(H_2(\{SK_{i-k}\}_x \| T_3 \| A_1 \| P I D_{V_i} \| $ $ m ))$ and $\phi_b^{\prime}=\mathcal{M}(H_2(\{SK_{i-k}\}_y \| T_3 \| A_1 \|  P I D_{V_i} \| m ))$.
   \item Then, $R_k$ uses a two-factor authentication process, PHY-SIAM and PHY-FTM, in two binary hypothesis testing problems for identity and message verification. This process comprises the following steps.\vspace{0.2cm}\\
    - \textit{Message verification step using PHY-SIAM}: In this step, $R_k$ uses the computed $\phi_a^{\prime}$ and $\phi_b^{\prime}$ to equalise the received PHY-layer signature in (\ref{e3}) by computing $\mathbf{R}_{1}^\prime=\mathbf{R}_{1} \odot {\mathbf{\Phi}_a^{\prime *}}$ and $\mathbf{R}_{2}^\prime=\mathbf{R}_{2} \odot {\mathbf{\Phi}_b^{\prime *}}$, where $\mathbf{\Phi}_a^\prime=$ $[e^{j{\phi_{a,1}^\prime}}, ..., e^{j{\phi_{a,N_2}^\prime}}]^T \in \mathbb{C}^{N_2 \times 1}$ and $\mathbf{\Phi}_b^\prime=$ $[e^{j{\phi_{b,1}^\prime}}, ...,$ $ e^{j{\phi_{b,N_2}^\prime}}]^T \in \mathbb{C}^{N_2 \times 1}$. Since $\xi_{1,i}$ and $\xi_{1,i}^{\prime}$ are highly correlated within $T_c$, $R_k$ verifies the received message by computing $\mathbf{C}={[c_1, ..., c_{N_2}]}^T=\mathbf{R}^\prime_{1} \odot \mathbf{R}^{\prime *}_{2}$. Then, $R_k$ calculates the circular variance $c.var(.)$ of $\angle (\mathbf{C})={[\angle (c_1), ..., \angle (c_{N_2})]}^T$ as
\begin{equation}\label{e4}
v={c.var}\left(\sum_{i=1}^{N_2} \arctan \left(\frac{\operatorname{Im}\left(c_i(t)\right)}{\operatorname{Re}\left(c_i(t)\right)}\right)\right)
\end{equation}
    where $c.var$ is defined as
\begin{equation}\label{e5}
\begin{gathered}
\alpha_i=\left(\begin{array}{c}
\cos \left(\angle\left(c_i\right)\right) \\
\sin \left(\angle\left(c_i\right)\right)
\end{array}\right), \bar{\alpha}=\frac{1}{N_2} \sum_{i=1}^{N_2} \alpha_i, \\
v=1-\|\bar{\alpha}\|
\end{gathered}
\end{equation}
    where $\|.\|$ represents the norm function. Avoiding impersonation and modification attacks, $R_k$ verifies $\sigma_{V_i}^{P H Y}$ in a hypothesis-testing problem given by 
\begin{equation}\label {e6}
\begin{gathered}
H_{0} \\
v \lessgtr \tau_{1} \\
H_{1}
\end{gathered}, \text { for } \begin{cases}H_{0}: & {\mathbf{\Phi}_a^\prime}={\mathbf{\Phi}_a} \hspace{0.1cm} \& \hspace{0.1cm} {\mathbf{\Phi}_b^\prime}={\mathbf{\Phi}_b} \\
H_{1}: & {\mathbf{\Phi}_a^\prime} \neq {\mathbf{\Phi}_a} \hspace{0.1cm} \& \hspace{0.1cm} {\mathbf{\Phi}_b^\prime} \neq {\mathbf{\Phi}_b}\end{cases}
\end{equation}
    where $\tau_{1}$ is the threshold value and $H_{0}$ and $H_{1}$ are the hypotheses that state whether the received message has been successfully authenticated or unauthenticated, respectively. For more information, see reference \cite{b1}.\vspace{0.1cm}\\
    - \textit{Message verification step using PHY-FTM}: Based on the OFDM symbols structure of order $M$ symbols in Fig. \ref{f4}, $R_k$ measures the correlation coefficient between the channel observation vector $\bar{H}_j$ estimated from the reference symbols of the $j^{th}$ OFDM symbol and that $\bar{H}_{j+1}$ of the $(j+1)^{th}$ OFDM symbol, starting from $\sigma_{V_i}^{P H Y}$ at $j=\{1, 2\}$ to the $M^{th}$ symbol. Hence, if $\bar{H}_j$ is highly correlated with $\bar{H}_{j+1}$, this means that these symbols are sent from the same transmitter. Otherwise, the received message is discarded. Hence, message verification can be described as a hypothesis-testing process based on the normalised likelihood ratio test (LRT), which is given by
        \begin{figure*}
\centerline{\includegraphics[width=17.8cm]{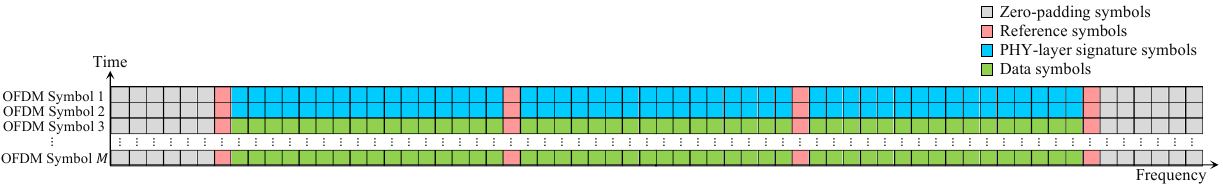}}
\setlength\belowcaptionskip{-0.5cm}
\caption{OFDM symbols' structure for 64 subcarriers.}
\label{f4}
\end{figure*}
    \begin{equation}\label{e7}
\begin{gathered}
\Lambda_{L R T}=\frac{n_{\tau_{2}}\left\|\bar{H}_{j}-\bar{H}_{j-1}\right\|^{2}}{\left\|\bar{H}_{j-1}\right\|^{2}} \hspace{0.3cm} \forall j \in [2, M], \hspace{0.3cm} \\
\hspace{0.5cm} H_{1} \\
\Lambda_{L R T}\lessgtr \tau_{2} \\
\hspace{0.5cm} H_{0}
\end{gathered}
\end{equation}
    where $\tau_{2} \in [0, 1]$ is the threshold value and $n_{\tau_{2}}$ is the normalisation coefficient. The decision rule can be made based on the sequential probability ratio test (SPRT) that sums the LRTs between the $j^{th}$ and the $(j-1)^{t h}$ OFDM symbols $\forall j \in[2, M]$. The SPRT-based hypothesis-testing problem can be expressed as 
    \begin{equation}\label{e8}
\begin{gathered}
\Lambda_{j}=\frac{n_{\tau_{2}}\left\|\bar{H}_{M-j+1}-\bar{H}_{M-j}\right\|^{2}}{\left\|\bar{H}_{M-j}\right\|^{2}} \hspace{0.05cm} \forall j \in [1, M-1],\\\Lambda_{S P R T}=n_{\tau_{3}} \sum_{j=2}^{M} \Lambda_{j}, \begin{gathered}
\hspace{0.65cm} H_{1} \\
\Lambda_{S P R T} \lessgtr \tau_{3} \\
\hspace{0.65cm} H_{0}
\end{gathered}
\end{gathered}
\end{equation}
    where $\tau_{3} \in [0, 1]$ is the threshold value and $n_{\tau_{3}}$ is the normalisation coefficient. For more information, see reference \cite{b1}.
\item Finally, $R_k$ accepts or discards the received message from $V_i$ based on the decision rule of both PHY-SIAM and PHY-FTM hypothesis problems. Accepted messages are those that are identified by both problems as being $H_0$. Otherwise, the message will be discarded.
\end{itemize}

\begin{figure}[t!]
\centerline{\includegraphics[width=9cm]{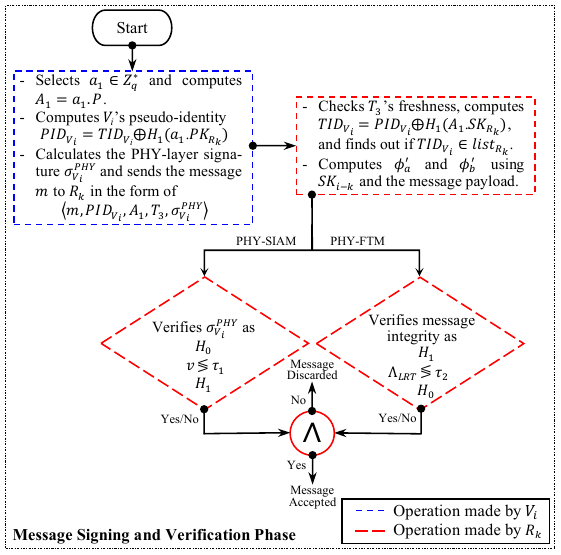}}
\setlength\belowcaptionskip{-0.5cm}
\caption{The top-level description of the message authentication and integrity verification phase.}
\label{f3}
\end{figure}

Fig. \ref{f3} shows the top-level description flowchart of the message authentication and integrity verification phase. Note that the pseudo-identity $P I D_{V_i}$ undergoes periodic updates for every re-authentication session to support traceability and anonymity. However, if we consider the system failure due to synchronization issues during the session time period denoted by $Q \times [100, 300] \hspace{0.1cm} msec$, the scheme will be reset and return to the initial authentication phase.

\subsection{RIS-assisted PHY-layer authentication}
One of the challenging issues of PHY-layer authentication is that the detection probability $P_d$ primarily depends on the received signal's SNR value, whereas $P_d$ defines the probability of authenticating legitimate users as authorised terminals. A higher SNR value indicates a higher $P_d$ for an acceptable false alarm probability $P_{fa}$, and vice versa, where $P_{fa}$ defines the probability of authenticating legitimate users as unauthorised terminals. This makes the PHY-layer authentication impractical in long-range and non-line-of-sight (NLoS) vehicular communications. While traditional beamforming and repeaters improve signal strength, their performance degrades in highly dynamic, obstructed vehicular environments due to factors like noise amplification and limited spectral efficiency. RIS offers a power-efficient, environment-aware solution that enhances throughput and robustness, especially under mobility and blockage conditions common in urban settings, see Fig. \ref{f1}. As a result, the proposed scheme can effectively authenticate the received messages from the vehicles in the shadowing areas. Thus, the received signals in (\ref{e3}) for the $i^{th}$ subcarrier is the superposition of $L$ multipath components coming from $L$ RIS's reflective elements and can be reformulated as 
\begin{equation}\label{e9}
\begin{aligned}
& r_{1,i}=(\mathbf{H}_{VI} \odot \mathbf{H}_{IR})\hspace{0.1cm} \boldsymbol{\omega}_\theta \hspace{0.1cm} s_{1,i}+n_i, \\
& r_{2,i}=(\mathbf{H}_{VI}^\prime \odot \mathbf{H}_{IR}^\prime)\hspace{0.1cm} \boldsymbol{\omega}_\theta \hspace{0.1cm} s_{2,i}+n_i^\prime
\end{aligned} \hspace{0.3cm} \forall i\in[1, N_2]
\end{equation}
where $\mathbf{H}_{VI}=[|h_{2,1}|e^{j \xi_{2,1}}, ..., |h_{2,L}|e^{j \xi_{2,L}}] \in \mathbb{C}^{1 \times L}$, $\mathbf{H}_{IR}=[|h_{3,1}|e^{j \xi_{3,1}}, ..., |h_{3,L}|e^{j \xi_{3,L}}] \in \mathbb{C}^{1 \times L}$, and $\boldsymbol{\omega}_\theta=[e^{j{\omega_1 \theta_1}}, ..., e^{j{\omega_L \theta_L}}]^T \in \mathbb{C}^{L \times 1}$. $\mathbf{H}_{VI}$ and $\mathbf{H}_{IR}$ represents the channel responses from $V_i$ to RIS and from RIS to $R_k$, respectively. While $\boldsymbol{\omega}_\theta$ defines the phase shift matrix related to the $L$ reflective elements of the RIS, where $\theta_l$ and $\omega_l$ defines the $l^{th}$ reflective element phase shift value and state, respectively $\forall l \in [1, L]$, for example,  $\theta_l=\pi$ and $\omega_l \in \{0, 1\}$ for a 1-bit RIS. Note that $\{\mathbf{H}_{VI}, \mathbf{H}_{IR}\}$ is highly correlated with $\{\mathbf{H}_{VI}^\prime, \mathbf{H}_{IR}^\prime\}$ within $T_c$. The RSU in each region optimises the RIS configuration $\boldsymbol{\omega}_\theta$ to maximise the power of the received signals at the side of the intended user. Hence, improving the receiver operating characteristics (ROCs; $P_d$ versus $P_{fa}$) of the two-factor re-authentication process at poor SNRs.

In conclusion, the impact of the RIS on the performance of the PHY-SIAM and PHY-FTM mechanisms can be summarised as follows:
\begin{enumerate}
    \item For PHY-SIAM: In the circular variance test defined in equations (\ref{e4} : \ref{e6}), the phase stability of the signal components \( \phi_{a,i}^\prime \) and \( \phi_{b,i}^\prime \) directly influences the dispersion of the resulting complex correlation vector \( \mathbf{C} \). Specifically, when the RIS is configured to align signal phases constructively, it preserves the coherence of the received signals originating from legitimate users. This coherence reduces the circular variance \( v \) under hypothesis \( H_0 \), resulting in a lower variance distribution of phase angles around the mean value-i.e., stronger clustering on the unit circle.

    \item \textbf{For PHY-FTM}: The RIS can be optimally configured to reinforce the dominant multipath component between a legitimate vehicle and the RSU. This reinforcement reduces the variability between successive channel estimates \( \bar{H}_{j-1} \) and \( \bar{H}_j \) in equations (\ref{e7}) and (\ref{e8}), thereby decreasing the term \( \left\|\bar{H}_j - \bar{H}_{j-1}\right\|^2 \) under hypothesis \( H_1 \) during normal (authentic) conditions.
\end{enumerate}

\section{Security and privacy analyses}\label{S4}
This section investigates how the proposed scheme satisfies the security and privacy requirements of VANETs.
\subsection{Attack model}
In this work, the attack model is designed to thoroughly assess the security of the proposed re-authentication method within VANET communications. The adversaries in this model are strategically positioned within the network to exploit various vulnerabilities: Eavesdroppers are located where they can intercept and potentially exploit sensitive communications between vehicles and RSUs. Replay attackers capture authentication messages and retransmit them to deceive the system into accepting outdated credentials. Impersonators forge credentials to appear as legitimate entities, gaining unauthorised access to the network. MitM attackers position themselves to intercept and alter communications between vehicles and RSUs, compromising the integrity of authentication messages. This attack model helps evaluate how well the proposed system withstands various types of security threats and maintains robustness under different adversarial conditions.
\subsection{Security proof using Random Oracle Modelling}
In cryptography, the random oracle model (ROM) is a theoretical framework often used to analyse the security of cryptographic constructions \cite{b37}. It involves using a random oracle, a mathematical function that produces random output for each unique input and maintains no internal state. In this part, we prove the security robustness of the proposed scheme using the ROM analysis. Specifically, we analyse the resistance of the signature generation process against potential threats posed by an adversarial entity $\mathcal{A}$. $\mathcal{A}$ is trying to impersonate an authorised vehicle $V_i$ by generating valid signatures $\sigma_{V_i}$ and $\sigma_{V_i}^{P H Y}$ within the initial authentication and message signing and verification phases. The computational complexity associated with signature generation in both phases depends on the infeasibility of forging two distinct cryptographic problems, formally defined as follows.

\begin{itemize}
        \item \textbf{Definition 1.} The elliptic curve discrete logarithm problem (ECDLP). Given $P P s$ and $Q = \gamma \cdot P$ on an elliptic curve, find $\gamma \in \mathbb{Z}_q^*$.
        \item\textbf{Definition 2.} Hashing problem. Given the value $s^{\prime}$, where $s^{\prime} = H_2(x)$, determine the corresponding input value $x \in \{0,1\}^{2 N_2}$.
\end{itemize}

The signature generation stage, denoted as $(q_s, q_k, \epsilon_{\text{Sig.Gen}})$, exhibits existential unforgeability against identity and adaptive chosen message attacks in the ROM, given that:

\begin{equation}\label{e50}
\epsilon_{\text {Sig.Gen }}=\epsilon\left(1-\frac{q_{s}^2 q_k^2}{q}\right)
\end{equation}
where, $q_s$ and $q_k$ represent the number of queries made to the oracles $\sigma_{V_i}(.)$ and $SK_{i-k}(.)$ respectively. Furthermore, $\epsilon_{\text{Sig.Gen}}$ denotes the probability of $\mathcal{A}$ to successfully generate a non-trivial forgery. The proof of (\ref{e50}) is given in Appendix \ref{A1}.

\subsection{Security proof using automated validation of internet security protocols and applications (AVISPA) simulation}
Similar to the work introduced in \cite{b31, b32}, this subsection proves the resilience of the proposed scheme against common attacks using the AVISPA simulation toolkit.
\begin{enumerate}
    \item \textit{Preliminaries}: Armando et al. \cite{b30} introduced the automated validation of internet security protocols and applications (AVISPA) framework, a prominent toolkit for assessing security protocols and internet applications. AVISPA employs high-level protocol specification language (HLPSL) to specify network terminal roles for agents, evaluating authentication and message confidentiality against potential intrusions. Security properties are formalized in ``goals,'' enabling protocol classification as ``SAFE'' or ``UNSAFE.'' The HLPSL2IF translator converts HLPSL code to the intermediate format (IF), feeding various back-ends: TA4SP, SATMC, OFMC, and CL-AtSe. In this study, CL-AtSe assesses the proposed scheme, evaluating resistance to MITM and replay attacks. A comprehensive explanation of these roles is offered within Appendix \ref{A2} through the utilisation of HLPSL codes. These codes are written using the notations and their corresponding AVISPA symbols listed in Table \ref{T2}.  It is important to note that the symbol ``$/\backslash$'' signifies a conjunction between two operations.
    \item \textit{Simulation specifications}: 
    The initial stage involves the definition of security objectives for the proposed scheme. These objectives encompass the authentication of broadcasted messages by the designated agent, as specified by auth\_1, auth\_2, and auth\_3. In the simulation context, two agent roles, namely role\_V1 and role\_RSU, are assumed, each representing the functions of V1 and RSU, respectively. The declarations pertinent to all agents are established within the role session, while the role environment identifies variables and functions associated with distinct agents. Fig. \ref{f3-} shows the protocol simulation, visualizing the transitional interactions among agents.

\begin{figure*}[t!]
\centerline{\includegraphics[width=15cm]{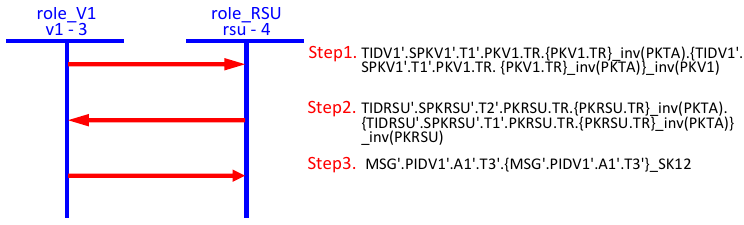}}
\setlength\belowcaptionskip{-0.5cm}
\caption{ Protocol simulation using AVISPA.}
\label{f3-}
\end{figure*}  
    \item \textit{Simulation results}:
    Using the AVISPA security analysis, simulation results of the specified goals are presented in Fig. \ref{f3--}, employing the Cl-AtSe back-end checker. Notably, the CL-AtSe model exhibits minimal time consumption ($\sim0.00$ seconds) for the IF translation process. Based on the summary, it can be inferred that the proposed scheme is secure against potential MITM and replay attacks.
    
\end{enumerate}

\subsection{Security and privacy informal analysis}
\subsubsection{Message authentication}
The proposed scheme offers legitimacy detection and ensures message integrity for the following reasons:

\begin{figure}[t!]
\centerline{\includegraphics[width=6cm]{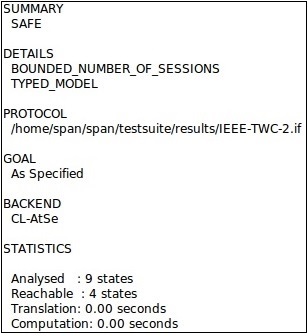}}
\setlength\belowcaptionskip{-0.5cm}
\caption{AVISPA simulation results using CL-AtSe.}
\label{f3--}
\end{figure}
\begin{itemize}
    \item For legitimacy detection, the recipient $V_i/R_k$ verifies the sender's legitimacy $R_k/V_i$ by checking if $Cert_{V_i/R_k} \in CRL$, where $\sigma_{TA} \in Cert_{V_i/R_k}$ is signed using $\beta \in Z_q^*$ and verified by the recipient using $PK_{TA} \in PPs$, which is infeasible to be forged under the difficulty of solving the ECDLP. In addition, the transmitted tuple $\langle TID_{V_i/R_k}, SPK_{V_i/R_k}, T_1, Cert_{V_i/R_k}, \sigma_{V_i/R_k} \rangle$ is verified for its integrity using the signature $\sigma_{V_i/R_k}$ that is signed using $V_i/R_k$'s secret key $Sk_{V_i/R_k}$ and verified by the recipient using $Pk_{V_i/R_k} \in Cert_{V_i/R_k}$.
    \item For message authentication at subsequent transmission slots, the tuple $\langle m, P I D_{V_i}, A_1, T_3,$ $ \sigma_{V_i}^{P H Y} \rangle$ is verified by $R_k$ for its integrity in a two-factor authentication process, PHY-SIAM and PHY-FTM, that's infeasible to be forged for the following reasons: A) The phase shifts, $\mathbf{\Phi}_a$ and $\mathbf{\Phi}_b$, in (\ref{e2}) are computed based on the shared key $SK_{i-k} \in \mathbb{G}$ and masked by $\mathbf{X}=\{e^{j{\psi_1}}, ..., e^{j{\psi_{N_2}}}\}$, where $\psi_i$ is a uniformly distributed random variable $ \sim U[0,2 \pi)$, which makes it infeasible for an adversary to differentiate between $\mathbf{\Phi}_a$ and $\mathbf{\Phi}_b$ and $\mathbf{X}$. B) The high correlation coefficient between subsequent channel observation vectors $\{\bar{H}_{j-1}, \bar{H}_j\}$ in (\ref{e7}) $\forall j \in [2, M]$ or $\{\bar{H}_j, \bar{H}_{j+1}\}$ in (\ref{e8}) $\forall j \in [1, M-1]$ helps in detecting modification attempts in the message payload.
\end{itemize}
\subsubsection{Privacy preservation}
In the proposed scheme, vehicles communicate using their temporary identities $TID_{V_i}$ at the first transmission slot, while pseudo identities $PID_{V_i}$ are used at subsequent transmissions. This preserves users' real identities $RID_{V_i}$ from exposure as no network terminals possess $RID_{V_i}$ or even the link between $RID_{V_i}$ and its associated long-term digital certificates $Cert_{V_i}$ except for the TA. Only the TA is authorised to expose $RID_{V_i}$ in cases of misbehaviour (for example, when the vehicle constructs an attack or when a driver drives an unregistered vehicle).
\subsubsection{Unlinkability}
For each $Q$ number of message transmissions per session, $V_i$ uses a different pseudo-identity $P I D_{V_i}=T I D_{V_i} \oplus H_1(a_1 . P K_{R_k})$, where $a_1 \in Z_q^*$ is dynamically updated for each session. Hence, no parameter is used twice per session, thereby avoiding location-tracking attacks.
\subsubsection{Traceability and revocation}
Each RSU in a specific area can report misbehaving vehicles to the TA by sending its associated digital certificate $Cert_{V_i}$. The TA, in turn, reveals its associated real identity, appends $Cert_{V_i}$ to the $CRL$, and distributes the updated $CRL$ among vehicles via RSUs.
\subsubsection{Perfect forward secrecy (PFS)}
PFS is a cryptographic property where the compromise of a long-term symmetric key does not compromise the confidentiality of past or future communications. It ensures that even if an attacker gains access to a session key, they cannot compromise previously recorded messages or intercept future messages ($m$). In this context, considering an adversary (Eve) capable of deducing the symmetric key $SK_{i-k}$ by hypothesising acquisition of the private key $SSK_{R_k}$ associated with RSU, the compromise of previous sessions becomes infeasible due to each session being established using a distinct randomly selected $S S K_{R_k} \in Z_q^*$. Consequently, no traceability exists among the established shared keys from different sessions.
\subsubsection{Resistance to passive and active attacks}
This part discusses the scheme's resistance against typical adversarial attacks. By considering an adversary, Eve acts as a passive attacker and listens to the communicating terminals' broadcasted messages to deduce any useful information about the symmetric key $Sk_{i-k}$. In this scenario, Eve attempts to deduce the shared key either during the initial authentication phase (case $1$) or during the message signing and verification phase (case $2$). In case $1$, $Sk_{i-k}$ is calculated using the Diffie-Hellman key exchanging protocol. This makes it difficult for Eve to compute $Sk_{i-k}$ due to the difficulty of solving the ECDLP. In case $2$, Eve has difficulty deducing the value of $Sk_{i-k}$ from the PHY-layer signature $\sigma_{V_i}^{P H Y}$ due to the following: $1)$ The signature generation step is dependent on the dynamically updated parameters $\langle T_i, A_{i}, PID_{V_i}, m\rangle$, which results in different outputs, $\mathbf{\Phi}_a$ and $\mathbf{\Phi}_b$, under the same shared key $Sk_{i-k}$. In addition, The received $\sigma_{V_i}^{P H Y}$ in (\ref{e3}) is dependent on the spatially and temporally varying channel phase responses $\xi_i$ and $\xi_{i}^{\prime}$ that masks $\phi_{a, i}$ and $\phi_{b, i}$, respectively. $2)$ For $y=H_2(x)$, it is difficult for Eve to determine the input variable $x$ from the hashed variable $y: \{0, 1\}^{N_2}$. In this scenario, we consider Eve to be an active attacker who is capable of constructing the following types of attacks:
\begin{itemize}
    \item \textit{Modification resistance}: In this attack, Eve tries to modify the message payload either during the initial authentication phase (case $1$) or during the message signing and verification phase (case $2$). In case $1$, the recipient $R_k/V_i$ verifies the received tuple $\langle TID_{V_i/R_k}, SPK_{V_i/R_k}, T_i,$ $ Cert_{V_i/R_k},$ $ \sigma_{V_i/R_k}\rangle$ for its integrity based on the attached signature $\sigma_{V_i/R_k}$. For this attack to be successful, Eve must modify the message contents and forge a valid signature, which is computationally intractable due to the difficulty of solving the ECDLP. In case $2$, Eve must modify the message contents $\langle m, P I D_{V_i}, A_i, T_i\rangle$ and forge a valid signature $\sigma_{V_i}^{P H Y}$. Without any knowledge of the shared key $Sk_{i-k}$, Eve is unable to correctly estimate the values of $\mathbf{\Phi}_a$ and $\mathbf{\Phi}_b$ needed to generate a valid signature. Accordingly, this type of attack can be easily detected.
    \item \textit{Impersonation resistance}: In this attack, Eve tries to impersonate the communicating vehicle $V_i$ during the initial authentication phase. For this attack to be successful, Eve must generate a valid signature $\sigma_{V_i}$ using the $V_i$'s secret key $Sk_{V_i}$, which cannot be forged due to the difficulty of solving the ECDLP. Accordingly, it is hard to compute a valid shared key $Sk_{i-k}$ used for generating $\sigma_{V_i}^{P H Y}$ during the message signing and verification phase. Hence, the proposed scheme is resistant to this type of attack.
    \item \textit{Replay resistance}: In this attack, Eve repeats the transmission of a previously captured message either during the initial authentication phase (case $1$) or during the message signing and verification phase (case $2$). In both cases, each transmission is accompanied by a fresh timestamp $T_i$ that helps the recipient detect this type of attack by testing whether $T_r-T_i\leq T_\Delta$ holds. Hence, the proposed scheme is resistant to replay attacks.
\end{itemize}
\subsection{Security proof using BAN-logic formal analysis}
The Burrows-Abadi-Needham (BAN) security proof is a formal methodology that offers a rigorous approach to evaluate the security of authentication protocols. The BAN approach is grounded in a formal model of authentication protocols and employs inference rules to analyse the knowledge and beliefs of principals involved in the protocol. Due to its effectiveness, the BAN methodology has been extensively adopted for analysing and verifying the security of authentication protocols in diverse settings such as computer networks, web communications, smart cards, and mobile devices. This study employs the BAN logic analysis to scrutinise the security of the proposed method against various types of attacks, such as replay, man-in-the-middle, and impersonation attacks.
\subsubsection{Notations}
In BAN-logic, security properties are expressed and argued using the following symbols.
\begin{itemize}
    \item $A \mid \equiv X$: $A$ believes that the proposition of $X$ is true.
    \item $A \triangleleft X$: $A$ sees $X$ denotes that principal $A$ has received a message that includes the value $X$.
    \item $A \mid \sim X$: $X$ has been transmitted to $A$ at some point, and $A$ has subsequently believed the proposition $X$.
    \item $A \mid \Longrightarrow X$: $A$ has control over the value $X$ and has the authority or jurisdiction to manipulate or modify it.
    \item $A \stackrel{k}{\longleftrightarrow} B$: $A$ and $B$ share a secret key $k$, which they use to securely communicate with each other.
    \item $A \stackrel{k}{\longrightarrow} B$: $k$ denotes the public key attributed to $A$.
    \item $\{X\}_{k}$: The shared key $k$ is used to encrypt $X$.
    \item $\# (X)$: It represents a fresh message $X$.
    \end{itemize}
\subsubsection{Rules}
A set of deductive rules are used to analyse initial beliefs and protocol messages exchanged between participants and make inferences about the security properties of the protocol. These rules are listed and defined in Table \ref{T3}.
\begin{table*}[t!]
\caption{The rules involved in the BAN-logic analysis}
\begin{center}
\setlength{\tabcolsep}{2pt} % Default value: 6pt
\begin{tabular}{l|l|l|l}
\hline \scriptsize\textbf{No.} & \scriptsize\textbf{Rule} & \scriptsize\textbf{BAN-logic representation} & \scriptsize\textbf{Definition} \\
\hline \scriptsize$R_1$ & \scriptsize Message rule for a shared key & \multicolumn{1}{c|}{\scriptsize$\frac{A \mid \equiv(A \stackrel{K}{\longleftrightarrow} B), A \triangleleft\{X\}_K}{A \mid \equiv(B \mid \sim X)}$} & \scriptsize If $A$ believes in $K$ and $A$ received $X$ encrypted by $K$, then $A$ believes $B$ said $X$ \\
\hline \scriptsize$R_2$ & \scriptsize Message rule for a public key & \multicolumn{1}{c|}{\scriptsize$\frac{A \mid \equiv(B \stackrel{K}{\longrightarrow} A), A \triangleleft\{X\}_{k^{-1}}}{A \mid \equiv(B \mid \sim X)}$} & \scriptsize If $A$ believes $K$ is $B$'s public key and receives $X$ encrypted with $B$'s private key,  \\
& & & \scriptsize then $A$ believes $B$ said $X$ \\
\hline \scriptsize $R_3$ & \scriptsize Nonce verification rule (NVR) & \multicolumn{1}{c|}{\scriptsize $\frac{A|\equiv \#(X), A| \equiv(B \mid \sim X)}{A \mid \equiv(B \mid \equiv X)}$} & \scriptsize If $A$ believes $X$ is fresh and that $B$ said $X$, then $A$ believes $B$ believes $X$ \\
\hline \scriptsize $R_4$ & \scriptsize Jurisdiction rule (JR) & \scriptsize $\frac{A|\equiv(B \Longrightarrow X), A| \equiv(B \mid \equiv X)}{A \mid \equiv X}$ & \scriptsize If $A$ believes $B$ has jurisdiction over $X$ and that $B$ believes $X$, then $A$ believes $X$ \\
\hline \scriptsize $R_5$ & \scriptsize Freshness rule (FR) & \multicolumn{1}{c|}{\scriptsize $\frac{A \mid \equiv \#(X)}{A \mid \equiv \#(X, Y)}$} & \scriptsize Freshness of one part ensures the freshness of the entire formula \\
\hline
\end{tabular}
\vspace{-0.5cm}\label{T3}
\end{center}
\end{table*}
\subsubsection{Goals}
The primary objective of BAN-logic is to demonstrate the validity of the proposed scheme by accomplishing the following set of goals.
\begin{itemize}
    \item \textit{Goal $1$}: $R_k \mid \equiv(R_k \stackrel{S k_{i-k}}{\longleftrightarrow} V_i)$.
    \item \textit{Goal $2$}: ${R_k \mid \equiv(V_i \mid \equiv M_1)}$.
    \item \textit{Goal $3$}: $V_i \mid \equiv(V_i \stackrel{S k_{i-k}}{\longleftrightarrow} R_k)$.
    \item \textit{Goal $4$}: ${V_i \mid \equiv(R_k \mid \equiv M_2)}$.
    \item \textit{Goal $5$}: $R_k \mid \equiv\left( M_3\right)$.
\end{itemize}
\subsubsection{Idealised forms}
The following points formulate the idealised messages for the proposed method.
\begin{itemize}
    \item $M_1$: $V_i \rightarrow R_k$: $\left\{{TID}_{V_i}, {SPK}_{V_i}, T_1, {Cert}_{V_i}\right\}_{S K_{V_i}}$, where ${Cert}_{V_i}=\{P K_{V_i}, T_R\}_{\beta}$.
    \item $M_2$: $R_k \rightarrow V_i$: $\left\{{TID}_{R_k}, {SPK}_{R_k}, T_2, {Cert}_{R_k}\right\}_{S K_{R_k}}$, where ${Cert}_{R_k}=\{P K_{R_k}, T_R\}_{\beta}$.
    \item $M_3$: $V_i \rightarrow R_k$: $\{m, P I D_{V_i}, A_1, T_3, \sigma_{V_i}^{P H Y}\}$, where $\sigma_{V_i}^{P H Y}=\{m, P I D_{V_i}, A_1, T_3\}_{S k_{i-k}}$.
\end{itemize}

\subsubsection{Assumptions}
The fundamental assumptions underlying the BAN-logic security proof are as follows.
\begin{itemize}
    \item $A_{1}$: $R_k \mid \equiv \#\left(T_{1}\right)$.
    \item $A_{2}$: $V_i \mid \equiv \#\left(T_{2}\right)$.
    \item $A_{3}$: $R_k \mid \equiv \#\left(T_{3}\right)$.
    \item $A_{4}$: $R_k \mid \equiv(T A \stackrel{K_{T A}}{\longrightarrow} R_k)$.
    \item $A_{5}$: $V_i \mid \equiv(T A \stackrel{K_{T A}}{\longrightarrow} V_i)$.
    \item $A_{6}$: $\frac{R_k \mid \equiv (TA \stackrel{PK_{TA}}{\longrightarrow} R_k), R_k \triangleleft\{Pk_{V_{i}}, T_{R}\}_{\beta}}{R_k \mid \equiv (V_i \stackrel{Pk_{V_{i}}}{\longrightarrow} R_k)}$.
    \item $A_{7}$: $\frac{V_i \mid \equiv (TA \stackrel{PK_{TA}}{\longrightarrow} V_i), V_i \triangleleft\{Pk_{R_k}, T_{R}\}_{\beta}}{V_i \mid \equiv (R_k \stackrel{Pk_{R_k}}{\longrightarrow} V_i)}$.
    \item $A_{8}$: $R_k|\equiv(V_i \Longrightarrow M_3)$.
\end{itemize}

\subsubsection{Implementation}
The security proof of the proposed protocol is presented as follows.
\begin{itemize}
    \item \textit{Step $1$}: Upon receipt of the message $M_1$ from $V_i$, $R_k$ applies $A_4$ and ${Cert}_{V_i} \in M_1$ to $A_6$, resulting in the following outcome: $O_1: {R_k \mid \equiv (V_i \stackrel{Pk_{V_{i}}}{\longrightarrow} R_k)}$. 
    \item \textit{Step $2$}: By applying $O_1$ and $M_1$ to $R_2$ from Table \ref{T3}, the outcome is  $O_2: {R_k \mid \equiv(V_i \mid \sim M_1)}$. Accordingly, $R_k$ computes $SK_{i-k}=SPK_{V_i}.SSK_{R_k}$ and have $O_3: R_k \mid \equiv(R_k \stackrel{SK_{i-k}}{\longleftrightarrow} V_i)$, achieving \textit{Goal $1$}. Next, by applying $A_1$ and $M_2$ to $R_5$ from Table \ref{T3}, we have $O_4: R_k \mid \equiv \#\left(M_{1}\right)$. Accordingly, by applying $O_4$ and $O_2$ to $R_3$ from Table \ref{T3}, we have ${R_k \mid \equiv(V_i \mid \equiv M_1)}$, achieving \textit{Goal $2$}.
    \item \textit{Step $3$}: Upon receipt of the message $M_2$ from $R_k$, $V_i$ applies $A_5$ and ${Cert}_{R_k} \in M_2$ to $A_7$, resulting in the following outcome: $O_5: {V_i \mid \equiv (R_k \stackrel{Pk_{R_k}}{\longrightarrow} V_i)}$. By applying $O_5$ and $M_2$ to $R_2$ from Table \ref{T3}, the outcome is  $O_6: {V_i \mid \equiv(R_k \mid \sim M_2)}$. Accordingly, $V_i$ computes $SK_{i-k}=SSK_{V_i}.SPK_{R_k}$ and have $O_7: V_i \mid \equiv(V_i \stackrel{SK_{i-k}}{\longleftrightarrow} R_k)$, achieving \textit{Goal $3$}.
    \item \textit{Step $4$}: By applying $A_2$ and $M_2$ to $R_5$ from Table \ref{T3}, we have $O_8: V_i \mid \equiv \#\left(M_{2}\right)$. Accordingly, by applying $O_8$ and $O_6$ to $R_3$ from Table \ref{T3}, we have ${V_i \mid \equiv(R_k \mid \equiv M_2)}$, achieving \textit{Goal $4$}. 
    \item \textit{Step $5$}: Upon receipt of the message $M_3$ from $V_i$, $R_k$ applies $O_2$ and $\sigma_{V_i}^{P H Y} \in M_3$ to $R_1$ from Table \ref{T3}, then we have $O_9: R_k \mid \equiv\left(V_i \mid \sim M_3\right)$. Next, by applying $A_3$ and $M_3$ to $R_5$ from Table \ref{T3}, we have $O_{10}: R_k \mid \equiv \#\left(M_{3}\right)$. Then, by applying $O_{10}$ and $O_{9}$ to $R_3$ from Table \ref{T3}, we have $O_{11}: R_k \mid \equiv\left(V_i \mid \equiv M_3\right)$. Finally, by applying $A_8$ and $O_{11}$ to $R_4$ from Table \ref{T3}, we have $O_{12}: R_k \mid \equiv\left( M_3\right)$, achieving \textit{Goal $5$}.
\end{itemize}
\section{Performance evaluation}\label{S5}
This section analyses the theoretical and practical aspects of RIS-assisted PHY-layer authentication performance, followed by detailed computation and communication comparisons.
\subsection{Theoretical analysis of the PHY-layer authentication}
\begin{figure}[t!]
\centerline{\includegraphics[width=9cm]{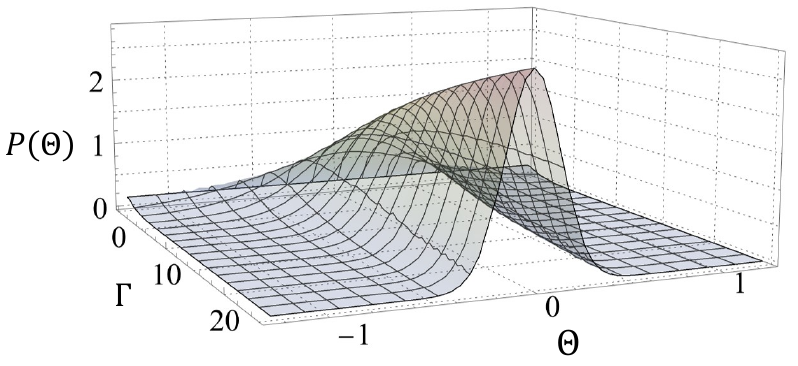}}
\setlength\belowcaptionskip{-0.5cm}
\caption{$P(\Theta \mid \Gamma)$ in (\ref{e10}) at different given $\Gamma \in [0, 25]$dB.}
\label{f10}
\end{figure}

In order to evaluate the ROCs of the proposed method, it is crucial to evaluate the probability density function (PDF) for the phase estimate ($\Theta$) of $\text { C }=\mathbf{R}_1^{\prime} \odot \mathbf{R}_2^{\prime *}$, where $\mathbf{R}_1^{\prime}$ and $\mathbf{R}_2^{\prime}$ denote the equalised received PHY-layer signature, given by the element-wise multiplication of $\mathbf{R}_1$ in (\ref{e3}) and $\boldsymbol{\Phi}_a^{\prime *}$, and $\mathbf{R}_2$ in (\ref{e3}) and $\boldsymbol{\Phi}_b^{\prime *}$, respectively. In the case of $\{\boldsymbol{\Phi}_a, \boldsymbol{\Phi}_b\}$ at the transmitting side $V_i$ are equivalent to $\{\boldsymbol{\Phi}_a^{\prime}, \boldsymbol{\Phi}_b^{\prime}\}$ at the receiving side $R_k$, the phase distribution of $\text { C }$ for varying SNR values can be formulated according to \cite{b1} as follows.
\begin{equation}\label {e10}
\begin{aligned}
P(\Theta \mid \Gamma)=& \frac{1}{2 \pi} e^{-\Gamma}+\frac{1}{\sqrt{\pi}}(\sqrt{\Gamma} \cos \Theta) \cdot\\
&  e^{-\Gamma \sin ^{2} \Theta}[1-\mathbb{Q}(\sqrt{2 \Gamma} \cos \Theta)]
\end{aligned}
\end{equation}
where
\begin{equation}\label {e11}
\begin{gathered}
\hspace{1.1cm}\Gamma=\frac{{|h_{i}|}^{2} \cdot {E_{S}}^2}{\sigma_n^2}, \\
\hspace{1.1cm}\mathbb{Q}(x)=\frac{1}{\sqrt{2 \pi}} \int_{x}^{\infty} e^{-t^{2} / 2} d t
\end{gathered}
\end{equation}
where $E_{S}$ is the symbol energy. Fig. \ref{f10} presents $P(\Theta \mid \Gamma)$ for different SNR values (i.e., $\Gamma \in [0, 25]$ dB). As indicated in (\ref{e4}), the circular variance of $\angle(\mathbf{C})$ with a specific order of $N_2$ is denoted as $v$, and this quantity satisfies the central limit theorem (CLT). Therefore, $v$'s distribution $\mathcal{F}(x)$ follows a normal distribution with a mean ($\mu_{H_0}$) equal to the variance of $P(\Theta)$ for a given $\Gamma$ value and a variance equal to $\sigma_{H_0}^2$. Thus, the following formulation can express $v$'s cumulative distribution function (CDF) for both hypotheses.
\begin{equation}\label {e12}
\phi\left(x \mid \mu_{H_{i}}, \sigma_{H_{i}}^{2}\right)=\frac{1}{2}\left[1+\operatorname{erf}\left(\frac{x-\mu_{H_{i}}}{\sqrt{2 \sigma_{H_{i}}^{2}}}\right)\right], \forall i \in \{0, 1\}
\end{equation}
In this context, we define $P_d=\left.\phi(x \mid{\mu_{H_{0}}, \sigma_{H_{0}}^{2}})\right|_{x=\tau_{1}}$ and $P_{fa}=\left.\phi(x \mid{\mu_{H_{1}}, \sigma_{H_{1}}^{2}})\right|_{x=\tau_{1}}$ for a threshold value $\tau_{1}$ of the hypothesis testing problem in (\ref{e6}). As illustrated in (\ref{e11}), the channel fading coefficient, represented by ${|h_{i}|}$, is a critical factor in determining the value of $\Gamma$ while maintaining a constant value of $E_s$ and noise variance $\sigma_n^2$. Generally, the received signal at the recipient side comprises various multipath components originating from distinct scatterers. Nonetheless, in this study, our focus is solely on the RIS path connecting the communicating terminals, as the impact of the remaining scatterers is consistent regardless of whether the RIS is being switched ON or OFF. The channel components of the $i^{th}$ subcarrier in both scenarios, considering the RIS turned ON and OFF, have been expressed in (\ref{e3}) and (\ref{e9}), respectively. Accordingly, the presence of the RIS can improve the SNR towards the communicating vehicle by configuring the reflective elements in a way that constructively interferes in a specific direction. This can be achieved by controlling the RIS electromagnetic behaviour by optimising $\boldsymbol{\omega}_\theta$ in (\ref{e9}) to maximise the $\Gamma$ value in (\ref{e11}). By doing so, the system's performance at a certain SNR value, denoted as $\Gamma = X$ dB, without the RIS can be equal to its performance at a lower SNR value, $\Gamma = X- \Delta X$ dB, with the RIS. A higher $\Gamma$ value signifies a decrease in the overlapping between the distributions of both hypotheses, $\mathcal{F}(x)|_{H_0}$ and $\mathcal{F}(x)|_{H_1}$, due to a lower value of $\mu_{H_0}$ for $\mathcal{F}(x)|_{H_0}$ relative to $\mu_{H_1}$ for $\mathcal{F}(x)|_{H_1}$. This improvement enhances the detection performance while maintaining an acceptable false alarm probability ($a_1$). Hence, the optimisation of the system's threshold value ($\tau_1$ in (\ref{e6})) can be computed by utilising the following formula \cite{b1}.
\begin{equation}\label{e13}
\tau_1=\arg \max _{\tau_1^{\prime}} \operatorname{erf}\left(\frac{\tau_1^{\prime}-\mu_{H_1}}{\sqrt{2 \sigma_{H_1}^2}}\right) \leq 2 a_1-1
\end{equation}
\begin{figure}[t!]
\centerline{\includegraphics[width=9.1cm]{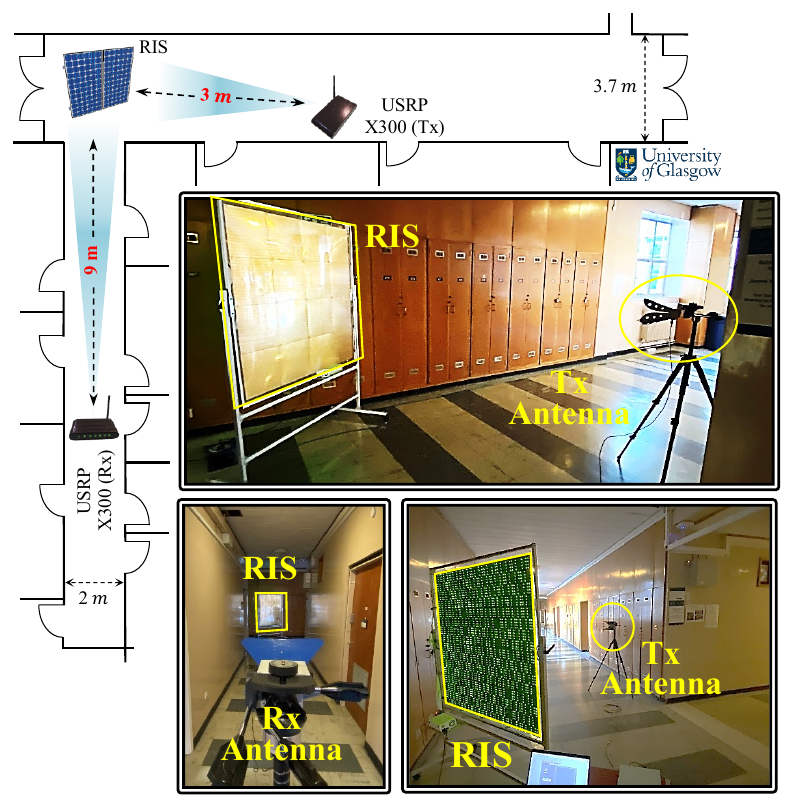}}
\caption{Experiment setup of the RIS-assisted method.}
\label{f5}
\end{figure}
\subsection{Practical experimentation of the RIS-assisted method}
In order to demonstrate the practicality of the proposed RIS-assisted PHY-layer authentication method, we conducted a hardware implementation using a $1$-bit RIS consisting of $4096$ reflective elements arranged in a two-dimensional $64 \times 64$ grid, along with a universal serial radio peripheral (USRP) equipped with two channels (denoted as $Ch_0$ and $Ch_1$) that functioned as the transmitter ($T_x$) and receiver ($R_x$), representing $R_k$ and $V_i$, respectively. The antennas used for $T_x$ and $R_x$ are of the two-horn type, with the $T_x$ antenna beam adjusted perpendicular to the RIS reflecting surface and located $3$ meters away from the centre. On the other hand, the $R_x$ antenna was situated $9$ meters away from the RIS, with an NLoS path between it and the $T_x$ antenna, and its beam set at a $45$-degree angle from the line connecting the $T_x$ antenna to the RIS. The experiment is conducted on an Intel Core $i7$ $2.7 \hspace{0.1 cm} GHz$ processor with 16.0 GB RAM with the NI LabVIEW platform to control the USRPs. Different views of the experimental setup are presented in Fig. \ref{f5}. Table \ref{T4} shows the experimental settings.

We set the carrier frequency $F_c$ to $3.75 \hspace{0.1 cm} GHz$ for 5G-V2I communication, the $T_x$ and $R_x$ gains to $20$ dB and $5$ dB, respectively, and the sampling rates for both channels to $200 \hspace{0.1cm} KHz$. We implemented a range of OFDM systems with varying numbers of subcarriers including $64$, $128$, and $256$, and cyclic prefix ($CP$) lengths of $16$, $32$, and $64$. To determine the optimal configuration associated with the location of the receiving antenna, we utilised the Hadamard codebook. The Hadamard codebook comprises a number of Hadamard matrices that provide a set of binary and orthogonal phase shift states ($\omega_l, \forall l \in [1, L]$) that can be used to modify the reflection of incoming electromagnetic waves in a desired direction or with a preferred phase shift by applying these values to the reflective elements, maximizing the SNR towards the $Rx$ side. Algorithm~(\ref{AA}) outlines the optimization process for selecting the optimal RIS configuration to maximize signal strength at the designated receiver. Accordingly, we implemented the proposed re-authentication method by transmitting two consecutive OFDM symbols with the same structure presented in Fig. \ref{f4}, representing the PHY-layer signature $\sigma_{V_i}^{P H Y}$.

\begin{figure}[t!]
\centerline{\includegraphics[width=6.5cm]{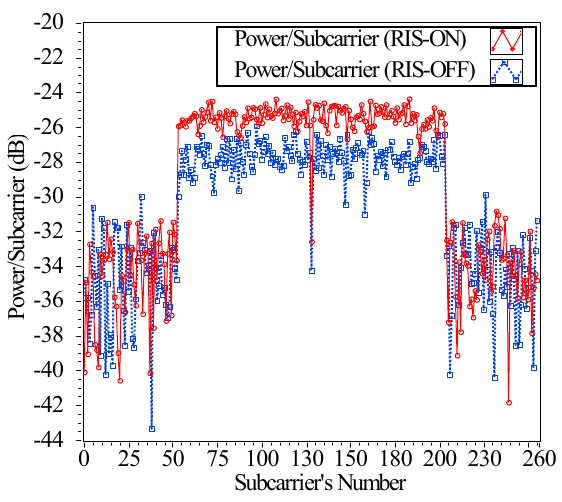}}
\setlength\belowcaptionskip{-0.5cm}
\caption{The received symbol's power for each subcarrier at $N=256$ subcarriers.}
\label{f9}
\end{figure}

\begin{figure}[t!]
\centerline{\includegraphics[width=6.5cm]{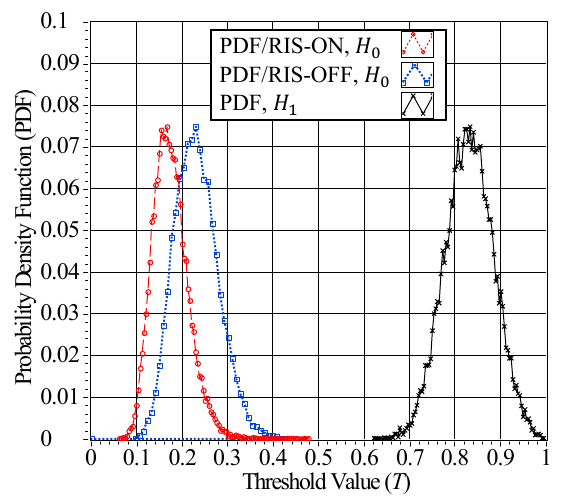}}
\setlength\belowcaptionskip{-0.5cm}
\caption{Distributions of both hypotheses $H_{0,1}$ with and without the RIS for $N=64$ subcarriers and SNR = $5$ dB.}
\label{f6}
\end{figure}

Fig. \ref{f9} shows the received OFDM symbol in the frequency domain after removing the $CP$ and applying the FFT. This figure presents the received power in dB for each subcarrier when the RIS is ON and OFF. Note that, the scenario of the deactivated RIS corresponds to the case described in reference [30]. It can be seen that the power of the subcarriers carrying data has increased by approximately $2$ dB with the activation of the RIS. This improvement is significant, especially for NLoS scenarios. Fig. \ref{f6} shows the PDF for hypothesis $H_0$ when the RIS is ON and OFF and for hypothesis $H_1$ for $N=64$ subcarriers and SNR = $5$ dB. The figure demonstrates that the activation of the RIS reduces the mean value for $\text{PDF}|_{H_0}$ compared to when the RIS is off. This reduction leads to a decrease in the overlap between $\text{PDF}|_{H_0}$ and $\text{PDF}|_{H_1}$, providing superior ROC curves under low SNR conditions.

\begin{algorithm}[t!] 
\footnotesize
\caption{Optimizing the Best RIS Configuration Towards Bob $(H_{opt})$}
\label{AA}
\begin{algorithmic}[1] 
\Statex \textbf{Initialization}
\State Construct the Hadamard codebook $HD = \{H_1, H_2, \ldots, H_L\}$ of $L$ matrices for the $(N_x \times N_y)$ RIS reflecting units
\State Initialize an empty list $SNR^{Bob}$ to store the SNRs measured at Bob

\Statex \textbf{SNR Measurement at the Legitimate Receiver (Bob)}
\For {$i = 1$ to $L$}
    \State Apply the Hadamard matrix configuration $H_i$ to the RIS
    \State Measure the average SNR $\overline{\mathrm{SNR}}_i^{Bob}$ at Bob
    \State Append $\overline{\mathrm{SNR}}_i^{Bob}$ to $SNR^{Bob}$
\EndFor

\Statex \textbf{Optimal Configuration Selection}
\State Find the index $i^* = \arg\max_i \left( \overline{\mathrm{SNR}}_i^{Bob} \right)$
\State Set the best RIS configuration: $H_{opt} = H_{i^*}$
\end{algorithmic}
\end{algorithm}
\vspace{0.3cm}

\begin{table}[!t]
\caption{Experimental settings}
\begin{center}
\setlength{\tabcolsep}{2pt} % Default value: 6pt
\begin{tabular}{l|l|l}
\hline \footnotesize	\textbf{Par.} & \footnotesize	\textbf{Value} & \footnotesize	\textbf{Description}\\
\hline $F_c$ & $3.75 \hspace{0.1 cm} GHz$ & Carrier frequency\\
\hline $T_x \hspace{0.1cm} (Gain)$ & $20$ dB & The transmitter gain\\
\hline $R_x \hspace{0.1cm} (Gain)$ & $5$ dB & The receiver gain \\
\hline $N$ & $64$, $128$, $256$ & Number of subcarriers\\
\hline $CP \hspace{0.1cm} length$ & $16$, $32$, $64$ & The cyclic prefix length\\
\hline $SR$ & $200 \hspace{0.1cm} KHz$ & The sampling rate for the $T_x$ and $R_x$\\
\hline $Antennas \hspace{0.1cm} types$ & Horn & $T_x$ and $R_x$ antennas types \\
\hline $T_x \leftrightarrow RIS$ & $3$ meters & The distance between the $T_x$ and RIS\\
\hline $RIS \leftrightarrow R_x$ & $9$ meters & The distance between the RIS and $R_x$\\
\hline
\end{tabular}
%\vspace{-0.8cm}
\label{T4}
\end{center}
\end{table}
\begin{figure*}[t!]
\centerline{\includegraphics[width=17cm]{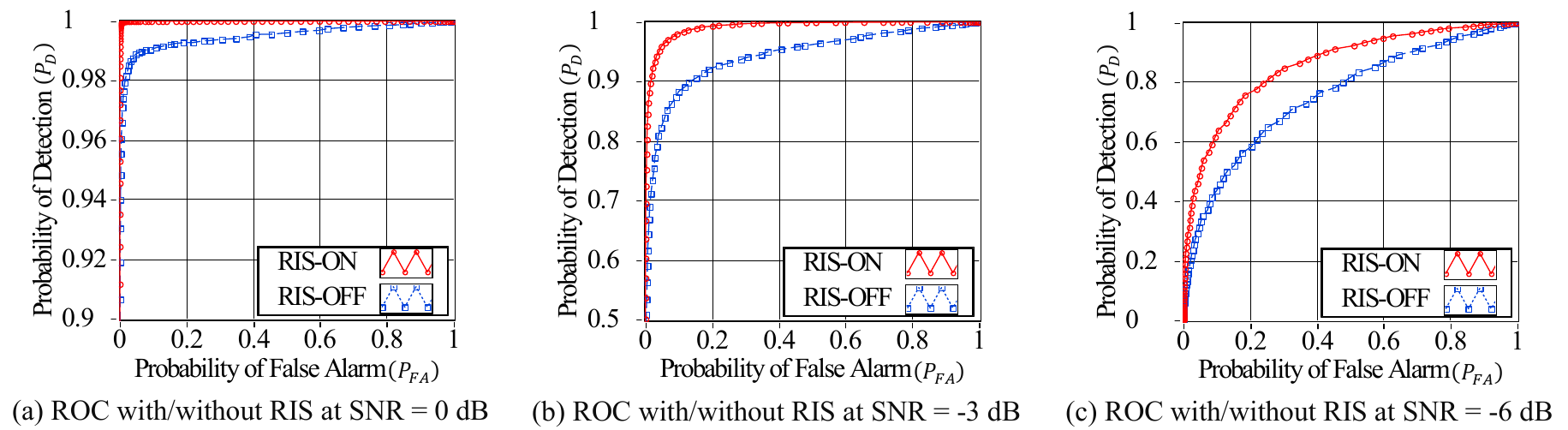}}
\setlength\belowcaptionskip{-0.2cm}
\caption{The ROCs with and without the RIS at different SNRs and $N=64$ subcarriers.}
\label{f7}
\end{figure*}

Fig. \ref{f7} illustrates the ROC curve for varying SNR values $\text{SNR}\in \{0, -3, -6\}$ dB, $N=64$ subcarriers, and with and without the use of the RIS. The figure demonstrates that decreasing the SNR value reduces $P_d$ for a given $P_{fa}$. This result arises from the increasing overlap between both hypotheses as the SNR decreases. Furthermore, the figure indicates that activating the RIS improves ROC curves. For example, when the RIS is off, the $P_d$ is approximately $0.92$; see Fig. \ref{f7}(b). However, with the RIS enabled, the $P_d$ increases to approximately $0.99$ for $P_{fa} \sim 0.2$, thereby demonstrating the ability of the RIS to enhance the authentication performance.

\begin{figure*}[t!]
\centerline{\includegraphics[width=17cm]{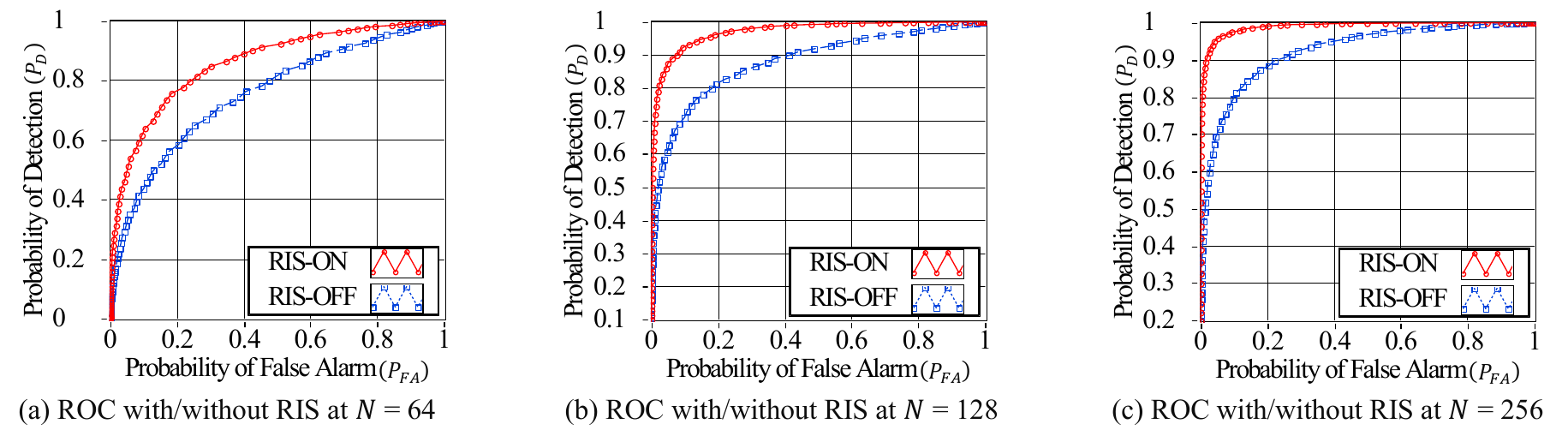}}
\setlength\belowcaptionskip{-0.1cm}
\caption{The ROCs with and without the RIS at different numbers of subcarriers and SNR $= -6$ dB.}
\label{f8}
\end{figure*}

Additionally, we evaluate the ROC for different numbers of subcarriers $N=\{64, 128, 256\}$ and fixed SNR value of $-6$ dB, as presented in Fig. \ref{f8}. Since $v$ in (\ref{e4}) represents the circular variance of a specific number of $N_2={3N/4}$ values, it follows the CLT. Hence, increasing the number of subcarriers results in an increase in $N_2$, which reduces the variance of $\mathcal{F}(x)|_{H_0}$ and minimises the overlap with $\mathcal{F}(x)|_{H_1}$, thereby improving the authentication performance. The enhanced ROC curves obtained in Fig. \ref{f8} affirm the effectiveness of increasing the number of subcarriers. Moreover, activating the RIS increases the $P_d$ for a given $P_{fa}$. As shown in Fig. \ref{f8}(b), when the RIS is off, the $P_d$ is roughly $0.82$. However, with the RIS enabled, the $P_d$ increases to approximately $0.96$ for $P_{fa} \sim 0.2$, thus demonstrating the beneficial impact of the RIS in enhancing authentication performance.

To further clarify the impact of mobility, it is important to note that the ROC of the proposed authentication scheme are influenced by two key factors: the $Tx$-$Rx$ separation distance, which determines the SNR, and the velocity of the moving $Rx$, which introduces Doppler components. In our prior work~\cite{b1}, we systematically investigated this relationship using a Doppler emulator at fixed SNR values in a realistic vehicular environment, demonstrating that mobility (i.e., varying speed) significantly shifts the ROC performance. However, the main objective of this study is to evaluate the role of RIS in reinforcing signal quality and improving re-authentication performance under challenging NLoS and low-SNR conditions. A full-scale evaluation of RIS reconfiguration under realistic mobility, while essential, requires adaptive optimization for each instantaneous $Tx/Rx$ position and is therefore considered for future work.

\subsection{Comparison of computation and communication costs}
This subsection presents the computation and communication analyses of the proposed method and shows that it outperforms traditional approaches. 

\subsubsection{Comparison of computation cost}
This part provides a detailed analysis of the computation comparison. Table \ref{T6} provides a summary of the running time for various crypto-based operations measured in \cite{b6} using the MIRACL cryptographic library \cite{b7} and a device equipped with an Intel Core $I7-6700$ processor. In Table \ref{T6}, the notations $\{T_{sm}^{BP}, T_{pa}^{BP}\}$ and $\{T_{sm}^{ECC}, T_{pa}^{ECC}\}$ denote the computational time for the BP-based and ECC-based scale multiplication and point addition, respectively. Furthermore, we evaluated the computational time for the mapping operation $T_{\mathcal{M}}$, and the circular variance operation in (\ref{e5}) denoted as $T_{c.var}$. The latter was insignificant compared to the values presented in Table \ref{T6}. Consequently, we have incorporated these results to accurately quantify the total computation cost of the proposed method and ensure a fair comparison, as listed in Table \ref{T5}.

\begin{table}[t!]
\caption{The time required for various crypto operations
in $msec$}
\begin{center}
\begin{tabular}{l|l|l}
\hline \textbf{Symbol} & \textbf{The operation definition} & \textbf{Run time} \\
\hline $T_{sm}^{BP}$ & BP-based scale multiplication in $\mathbb{G}_1$ & $0.6940$ \\
 $T_{pa}^{BP}$ & BP-based point addition in $\mathbb{G}_1$ & $0.0018$ \\
 $T_{sm}^{ECC}$ & ECC-based scale multiplication in $\mathbb{G}$ & $0.3218$\\
 $T_{pa}^{ECC}$ & ECC-based point addition in $\mathbb{G}$ & $0.0024$\\
 $T_h$ & One way hashing operation & $0.0010$\\
\hline
\end{tabular}
\vspace{-0.5cm}
\label{T6}
\end{center}
\end{table}
In our proposed scheme, the EC signature generation process incurs a cost of approximately $1 T_{sm}^{ECC}$, while the verification process costs $2 T_{sm}^{ECC}$. Based on this, the computation cost of transmitting $n$ messages from a single vehicle using our method can be expressed as $[3 T_{sm}^{ECC} + \lceil \frac{n}{Q} \rceil(2 T_{sm}^{ECC} + T_h)+n(T_h+T_{\mathcal{M}})]$. The first term accounts for the signature generation and the secret key agreement, the second term accounts for the dynamically updating pseudo-identity after every $Q$ transmitted messages, and the third term accounts for generating $\sigma_{V_i}^{PHY}$. On the other hand, the verification time can be expressed as $[2 T_{sm}^{ECC}+\lceil \frac{n}{Q} \rceil (T_{sm}^{ECC} + T_h)+n(T_h+T_{\mathcal{M}}+T_{c.var})]$. The first term corresponds to the initial signature verification, and the second and third terms verify the pseudo-identity for every $Q$ transmitted messages and $\sigma_{V_i}^{PHY}$, respectively. Thus, the total computation cost can be expressed as $(0.9654+0.3228\lceil\frac{n}{q}\rceil+0.001n) \hspace{0.1cm} msec$.
\begin{table*}[t!]
\caption{Computation and communication comparisons}
\begin{center}
\setlength{\tabcolsep}{2pt} % Default value: 6pt
\renewcommand{\arraystretch}{1.1} % Default value: 1
\begin{tabular}{c|c|c|c|c}
\hline \multirow{3}{*}{ \scriptsize\textbf{Scheme}} & \multicolumn{3}{|c|}{ \scriptsize\textbf{Computation cost (\textit{msec})}} & \multirow{1}{*}{\scriptsize\textbf{Communication }} \\
\cline { 2 - 4 } & \scriptsize\textbf{Signature generation} & \scriptsize\textbf{Signature verification of $n$ messages} & \scriptsize  \textbf{Verifying 10000 messages} & \scriptsize\textbf{cost (\textit{bytes})}\\
\hline \scriptsize  \cite{b120} & \scriptsize  $T_{sm}^{ECC} + T_{pa}^{ECC}$ & \scriptsize   $T_{sm}^{ECC} + (n) T_{pa}^{ECC}$ & \scriptsize 24.3218  &\scriptsize  $128n$ \\
\hline \scriptsize \cite{b3} & \scriptsize $3 T_{sm}^{ECC} + 3 T_h$ & \scriptsize $(n+2) T_{sm}^{ECC} + (n-1) T_{pa}^{ECC} + (2n) T_h $ & \scriptsize  3262.64 & \scriptsize $124n$ \\
\hline \scriptsize \cite{b4} & \scriptsize $2 T_{sm}^{BP} + 2 T_{pa}^{BP} + T_h$ & \scriptsize $(3n+2) T_{sm}^{BP} + (2n) T_{pa}^{BP} + (n) T_h$ & \scriptsize 20867.4 &\scriptsize $300n$ \\
\hline \cite{b5} & $3 T_{sm}^{BP}+2 T_{pa}^{BP}+ T_h$ & $(3 n+2) T_{sm}^{BP}+(3 n) T_{pa}^{BP}+(n) T_h$ & \scriptsize 20885.4 & $408n$ \\
\hline \scriptsize  \cite{b121} & \scriptsize  $2 T_{sm}^{ECC} + 2 T_{pa}^{ECC} + T_h$ & \scriptsize  $(n+1) T_{sm}^{ECC} + (3n-1) T_{pa}^{ECC} + (n) T_h$ & \scriptsize 3300.32 &\scriptsize  $124n$ \\
\hline \scriptsize Ours &\scriptsize $3 T_{sm}^{ECC} + \lceil \frac{n}{Q} \rceil(2 T_{sm}^{ECC} + T_h)+n(T_h+T_{\mathcal{M}})$ &\scriptsize $2 T_{sm}^{ECC}+\lceil \frac{n}{Q} \rceil (T_{sm}^{ECC} + T_h)+n(T_h+T_{\mathcal{M}}+T_{c.var})$ & \scriptsize 13.8716 &\scriptsize $188+112n$ \\
\hline
\end{tabular}
\vspace{-0.5cm}
\label{T5}
\end{center}
\end{table*}

\begin{figure}[t!]
\centerline{\includegraphics[width=8.5cm]{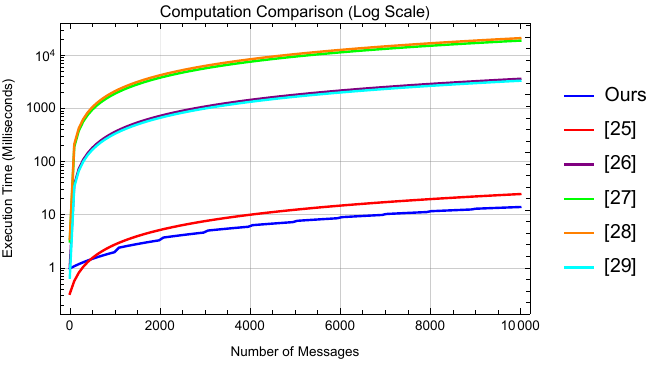}}
\setlength\belowcaptionskip{-0.5cm}
\caption{The computation cost of verifying $10000$ messages at $Q=1000$.}
\label{f11}
\end{figure}

In Mohammed et al. \cite{b120}, the computation cost for verifying $n$ received messages is $[T_{sm}^{ECC} + (n) T_{pa}^{ECC}] = (0.3218 + 0.0024n) \hspace{0.1cm} msec$, while for Cui et al. \cite{b3}, Wang et al. \cite{b4}, Li et al. \cite{b5}, and Almazroi et al. \cite{b121}, this value is $[(n+2) T_{sm}^{ECC} + (n-1) T_{pa}^{ECC} + (2n) T_h] = (0.6412+0.3262n) \hspace{0.1cm} msec$, $[(3n+2) T_{sm}^{BP} + (2n) T_{pa}^{BP} + (n) T_h] = (1.388+2.0866n) \hspace{0.1cm} msec$, $[(3 n+2) T_{sm}^{BP}+(3 n) T_{pa}^{BP}+(n) T_h] = (1.388 + 2.0884n) \hspace{0.1cm} msec$, and $[(n+1) T_{sm}^{ECC} + (3n-1) T_{pa}^{ECC} + (n) T_h] = (0.3194 + 0.33n) \hspace{0.1cm} msec$, respectively. To illustrate the comparison, Fig. \ref{f11} displays the computation cost required to verify $10000$ received messages from a single user. Table \ref{T5} and Fig. \ref{f11} indicate that the proposed scheme has a computational cost of approximately $13.87 \hspace{0.1 cm} \text{ms}$, which is nearly half that of its closest competitors reported in \cite{b120}.

\begin{figure}[t!]
\centerline{\includegraphics[width=8.5cm]{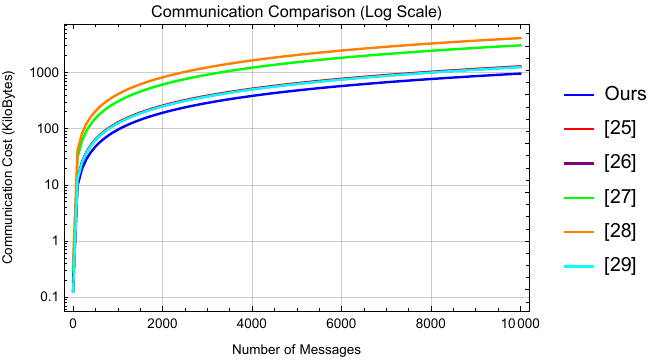}}
\setlength\belowcaptionskip{-0.5cm}
\caption{The communication cost of sending $10000$ messages.}
\label{f12}
\end{figure}

\subsubsection{Comparison of communication cost }
This part provides a detailed comparison of communication costs. For the $80$-bit security level of the proposed scheme, the elliptic curve group is denoted as $\mathbb{G}$, where $|\mathbb{G}|=40$ bytes and $Z_q^* = 20$ bytes. For the same security level, the bilinear pairing is denoted as $\bar{E}: \mathbb{G}_1 \times \mathbb{G}_1 \rightarrow \mathbb{G}_T$, where $\bar{P}$ is the generator of the elliptic curve $\bar{E}: y^2=x^3+x \bmod \bar{p}$, with $|\mathbb{G}_1|=128$ bytes and $Z_{\bar{q}}^* = 20$ bytes. Moreover, the size of hashed values using the SHA-$1$ hashing operation is $20$ bytes, and the timestamp has a size of $4$ bytes.

In the proposed scheme, the communication cost of transmitting $n$ messages is determined by the size of the tuple $\langle{TID}_{{V_i}}, SPK_{V_i}, T_1, (PK_{V_i}, T_R, \sigma_{TA}), \sigma_{V_i}\rangle$ during the first transmission slot, as well as the size of the tuple $\left\langle P I D_{V_i}, A_1, T_3, \sigma_{V_i}^{P H Y}\right\rangle$ for $n$ subsequent transmissions. Specifically, $\{SPK_{V_i}, PK_{V_i}, A_1\} \in \mathbb{G}$, and the length of ${TID}_{{V_i}}$ and ${PID}_{{V_i}}$ is $20$ bytes each. The sizes of $\sigma_{TA}$ and $\sigma_{V_i}$ are $40$ bytes each, while the lengths of $T_R$, $T_1$, and $T_3$ are $4$ bytes each. The size of $\sigma_{V_i}^{P H Y}$ is $48$ bytes. Therefore, the total communication cost for transmitting $n$ messages is $[(20+ 2 \times 4+4 \times 40)+(20+40+4+48)n]=(188+112n)$ bytes. In Mohammed et al. \cite{b120}, the signature is represented by the tuple $\langle P I D_i^1, P I D_i^2, U_i, R_i, T_i \rangle$, where $|P I D_i^1| = 64$ bytes, $\{P I D_i^2, U_i, R_i\} \in Z_q^*$ and $T_i$ is timestamp. Thus, the total size of the signature is $(64 + 3 \times 20 + 4) = 128$ bytes. In Cui et al. \cite{b3}, the signature is represented by the tuple $\langle P I D_{j}^1, P I D_{j}^2, \delta_j, D_j, T_j\rangle$, where $\{P I D_{j}^1, D_j\} \in \mathbb{G}$, $\{P I D_{j}^2, \delta_j\} \in Z_q^*$, and $T_j$ denotes the timestamp. Thus, the total size of the signature is $( 2 \times 40 + 2 \times 20 + 4 ) = 124$ bytes. In Wang et al. \cite{b4}, the signature is represented by the tuple $\langle R_{u_i}, T_{u_i}^{\prime}, \varrho_{u_i}, P K_{T A}, t_i\rangle$, where $\{R_{u_i}, T_{u_i}^{\prime}\} \in \mathbb{G}_1$, $\{\varrho_{u_i}, P K_{T A}\} \in Z_{\bar{q}}^*$, and $t_i$ represents the timestamp. Thus, the total size of the signature is $( 2 \times 128 + 2 \times 20 + 4 ) = 300$ bytes. Similarly, Li et al. \cite{b5} represent a signature as $\langle R_{u_i}, K_{u_i}^{\prime}, K G_{u_i}^{\prime}, \varrho_{u_i}, t_i \rangle$, where $\{R_{u_i}, K_{u_i}^{\prime}, K G_{u_i}^{\prime}\} \in \mathbb{G}_1$, $\varrho_{u_i} \in Z_{\bar{q}}^*$, and $t_i$ denotes the timestamp. The total size of this signature is $( 3 \times 128 + 20 + 4 ) = 408$ bytes.
In Almazroi et al. \cite{b121}, the signature is represented by the tuple $\langle A I D_v, R_{p u b}, D_{v, i}, T_{V, i}, \sigma_{v, i} \rangle$, where $\{R_{p u b}, D_{v, i}\} \in \mathbb{G}$, $\{A I D_v, \sigma_{v, i}\} \in Z_q^*$, and timestamp $T_{V, i}=4$ bytes. Thus, the total size of this signature is $(2 \times 20 + 2 \times 40 + 4) = 124$ bytes. Fig. \ref{f12} shows the communication cost required for transmitting $10000$ messages received from a single user. Our proposed scheme exhibits the lowest communication cost of at least $\approx 120 \hspace{0.1cm} Kb$ less than that of its best competitors in \cite{b3}.

\section{Conclusions}\label{S6}
This paper proposes an authentication scheme that utilises the RIS to enhance the detection probability of the PHY-layer authentication in NLoS conditions while still adhering to the security and privacy requirements of VANETs. The theoretical and experimental results demonstrate the effectiveness of the RIS in improving authentication performance. We have performed informal and formal (BAN-logic) analyses to verify the scheme's security resistance against typical attacks. Additionally, we have conducted a computation and communication comparison to demonstrate that the proposed method reduces overheads, resulting in a computation cost savings of over $98\%$ compared to existing methods in \cite{b3, b4, b5}, and communication cost savings of approximately $10\%$, $62\%$, and $72\%$ compared to \cite{b3}, \cite{b4}, and \cite{b5}, respectively. Our future work will explore the possibility of incorporating the PHY-layer secret key extraction as an alternative key agreement technique and implementing it in outdoor scenarios. Furthermore, we aim to extend the proposed scheme to dynamic vehicular environments by employing a mobile USRP testbed mounted on a railkit and integrating adaptive RIS configuration using machine learning to evaluate performance under realistic mobility conditions. 

\appendix 
\subsection{ROM security prove}\label{A1}
\textit{Proof}. Considering an adversary $\mathcal{A}$ who is trying to impersonate the vehicle $V_i$ during the initial authentication and message signing and verification phases. In this case, $\mathcal{A}$ must forge a valid signature $\sigma_{V_i}$ and $\sigma_{V_i}^{P H Y}$ by the construction of an algorithm $C$ to solve the defined problems with a probability of success $\epsilon_{\text {Sig.Gen.}}$. Algorithm $C$ initially holds two empty tables $T_{\sigma_{V_i}}[.]$ and $T_{SK_{i-k}}[.]$ to simulate random oracles $\sigma_{V_i}(.)$ and $SK_{i-k}(.)$, then answers $\mathcal{A}$'s oracle queries as follows:
\begin{itemize}
    \item \textit{$\sigma_{V_i}(.)$ queries}: For a query $\sigma_{V_i}^{\prime}$, $C$ holds $\{TID_{V_i}, Cert_{V_i}\}$, selects $\{S S K_{V_i}, S K_{V_i}\} \stackrel{\$}{\leftarrow} Z_q^*$ at $T_1$ timestamp, and calculates $S P K_{V_i}=S S K_{V_i} \cdot P$. Then, $C$ generates its associated EC-signature using $T_{\sigma_{V_i}}[.]$, so that $T_{\sigma_{V_i}}[SK_{V_i}, {TID_{V_i}, T_1, S P K_{V_i}, Cert_{V_i}}] {\leftarrow} \sigma_{V_i}^{\prime} \in \mathbb{G}$. of order $q$. If $T_{\sigma_{V_i}}[SK_{V_i}, \{TID_{V_i}, T_1, S P K_{V_i}, Cert_{V_i}\}]$ is defined, $C$ halts, returns $\perp$, and sets $false \leftarrow true$. Otherwise, it returns $\sigma_{V_i}^{\prime}$ to $\mathcal{A}$ under $SK_{V_i}$.
    \item \textit{$SK_{i-k}(.)$ queries}: For a query $SK_{i-k}^{\prime}$, $C$ holds $m$, randomly selects $a_1 \stackrel{\$}{\leftarrow} Z_q^*$, and computes $A_1=a_1 \cdot P$ and $P I D_{V_i}=T I D_{V_j} \oplus H_1\left(a_1 \cdot P K_{R_k}\right)$. Then, $C$ generates the shared key $SK_{i-k}$ using $T_{SK_{i-k}}[.]$, so that $T_{SK_{i-k}}[S S K_{V_i}] \leftarrow SK_{i-k }^{\prime} \in \mathbb{G}$ of order $q$. If $T_{SK_{i-k}}[S S K_{V_i}]$ is defined, $C$ halts, returns $\perp$, and sets $false \leftarrow true$. Otherwise, it returns $SK_{i-k}^{\prime}$ to $\mathcal{A}$ under $S S K_{V_i}$.
\end{itemize}
Finally, it is assumed that $\mathcal{A}$ successfully generated a forged signature $\langle{TID}_{V_i}, S P K_{V_i}, T_1, {Cert}_{V_i},$ $ \sigma_{V_i}^{\prime}\rangle$ under $SK_{V_i}$ and $\left\langle m, P I D_{V_i}, A_1, T_3, \sigma_{V_i}^{P H Y}\right\rangle$ under $S S K_{V_i}^{\prime}$ based on $q_{s}$ and $q_k$ queries for $\sigma_{V_i}(.)$ and $SK_{i-k}(.)$ oracles with probability $\epsilon_{\text {Sig.Gen }}=\operatorname{Pr}\left[E_1\right] \operatorname{Pr}\left[E_2 \mid E_1\right]$, in which $E_1$ and $E_2$ are defined as:
\begin{itemize}
    \item \textit{Event $E_1$} : The execution of Algorithm $C$ did not halt as a result of $\{\sigma_{V_i}^{\prime}, SK_{i-k}^{\prime}\}$ generation.
    \item \textit{Event $E_2$} : Adversary $\mathcal{A}$ successfully generates a non-trivial forgery.
\end{itemize}
The probability $\operatorname{Pr}[\neg \text{false}]$ is to be calculated, where $\neg \text{false}$ signifies that algorithm $C$ does not abort due to the $\sigma_{V_i}(.)$ and $SK_{i-k}(.)$ queries. This probability is assessed based on the claims outlined below.
Claim 1. $\operatorname{Pr}\left[E_1\right]=\operatorname{Pr}[\neg false ] \geq 1-\frac{q_{s}^2 q_k^2}{q^2}$\\
\textit{Proof}. The probability $\operatorname{Pr}[\text{false}]$ can be determined by approximating the product of the following probabilities.
\begin{itemize}
    \item \textit{Scenario 1}. The event $false \leftarrow true$ is realised during the $\sigma_{V_i}(.)$ queries when $\sigma_{V_i}^{\prime}$ previously recorded in $\sigma_{V_i}(.)$ oracle under $SK_{V_i}$. With a maximum of $q_{s}$ queries recorded in the $T_{\sigma_{V_i}}[.]$ table, the probability for an individual $\sigma_{V_i}(.)$ query remains at most $\frac{q_{s}}{q}$, and the cumulative probability for $q_{s}$ queries stands at $\frac{q_{s}^2}{q}$.
    \item \textit{Scenario 2}. The event $false \leftarrow true$ is realised during the $SK_{i-k}(.)$ queries when $SK_{i-k}^{\prime}$ previously recorded in $SK_{i-k}(.)$ oracle under $S S K_{V_i}$. With a maximum of $q_{k}$ queries recorded in the $T_{SK_{i-k}}[.]$ table, the probability for an individual $SK_{i-k}(.)$ query remains at most $\frac{q_{k}}{q}$, and the cumulative probability for $q_{k}$ queries stands at $\frac{q_{k}^2}{q}$.
\end{itemize}
Claim 2. $\operatorname{Pr}\left[E_2 \mid E_1\right] \geq \epsilon$\\
\textit{Proof}. The term $\operatorname{Pr}\left[E_2 \mid E_1\right]$ denotes the probability that adversary $\mathcal{A}$ generates an authentic forgery, and algorithm $C$ continues without termination due to the $\mathcal{A}$'s interactions involving $\sigma_{V_i}(.)$ and $SK_{i-k}(.)$ queries. This indicates that all responses to these queries are valid. As a result, adversary $\mathcal{A}$ is able to fabricate a valid forgery with a probability denoted by $\epsilon$. Consequently, the probability that adversary $\mathcal{A}$ effectively impersonate $V_i$ through the production of a significant forgery under the context of $\{SK_{V_i}, SSK_{V_i}\}$ is at least:
$$
\epsilon_{\text {Sig.Gen }}=\epsilon\left(1-\frac{q_{s}^2 q_k^2}{q}\right)
$$

\subsection{AVISPA simulation}\label{A2}
This subsection presents the AVISPA simulation codes.

\begin{table}[H]
\begin{center}
\setlength{\tabcolsep}{2pt} % Default value: 6pt
\renewcommand{\arraystretch}{0.6} % Default value: 1
\begin{tabular}{| p{8cm} |}
\hline 
\scriptsize\textbf{Code 1}\text{: HLPSL code for the role of the vehicle $V_{1}$, played by $V_{1}$}\\
\hline
\scriptsize\text{\textcolor{red}{role role\_V1} (V1,RSU:agent,PKTA,PKV1,PKRSU:public\_key,}\\
\hspace{1.45cm} \scriptsize\text{SK12:symmetric\_key,SND,RCV:channel(dy))}\\
\scriptsize\text{played\_by V1}\\
\scriptsize\text{def=}\\
\scriptsize\hspace{0.8cm} \textcolor{blue}{local}\\
\scriptsize\hspace{1.4cm} \text{State:nat,TR,T1,T2,T3,TIDV1,SPKV1,TIDRSU,}\\
\scriptsize\hspace{1.4cm} \text{SPKRSU,MSG,PIDV1,A1:text}\\
\scriptsize\hspace{0.8cm} \textcolor{blue}{init}\\
\scriptsize\hspace{1.4cm} \text{State:=0}\\
\scriptsize\hspace{0.8cm} \textcolor{blue}{transition}\\
\scriptsize\hspace{1.4cm} \text{1. State=0 $/\backslash$ RCV(start) $=|>$ State' := 1 $/\backslash$ T1' :=} \\ 
\scriptsize\hspace{1.7cm} \text{new() $/\backslash$ TIDV1' := new() $/\backslash$ SPKV1' := new() $/\backslash$}\\
\scriptsize\hspace{1.7cm} \text{SND(TIDV1'.SPKV1'.T1'.PKV1.TR.\{PKV1.TR\}}\\
\scriptsize\hspace{1.7cm} \text{\_inv(PKTA).\{TIDV1'.SPKV1'.T1'.PKV1.TR.\{PKV1.}\\
\scriptsize\hspace{1.7cm} \text{TR\}\_inv(PKTA)\}\_inv(PKV1))}\\
\scriptsize\hspace{1.7cm} \textcolor{teal}{\%\% V1 hopes that SPKV1' will be verified by RSU}\\
\scriptsize\hspace{1.7cm} \text{$/\backslash$ witness(V1,RSU,auth\_1,SPKV1')}\\
\vspace{-0.03cm}\scriptsize\hspace{1.4cm} \text{2. State=1 $/\backslash$ RCV(TIDRSU'.SPKRSU'.T2'.PKRSU.}\\
\scriptsize\hspace{1.7cm} \text{TR.\{PKRSU.TR\}\_inv(PKTA). \{TIDRSU'.}\\
\scriptsize\hspace{1.7cm} \text{SPKRSU'.T1'.PKRSU.TR.\{PKRSU.TR\}}\\
\scriptsize\hspace{1.7cm} \text{\_inv(PKTA)\}\_inv(PKRSU)\} $=|>$ State' := 2 $/\backslash$ }\\
\scriptsize\hspace{1.7cm} \text{T3' := new() $/\backslash$ PIDV1' := new() $/\backslash$ A1' := new() }\\
\scriptsize\hspace{1.7cm} \text{$/\backslash$ MSG' := new() $/\backslash$ SND(MSG'.PIDV1'.A1'.T3'.}\\
\scriptsize\hspace{1.7cm} \text{\{MSG'.PIDV1'.A1'.T3'\}\_SK12)}\\
\scriptsize\hspace{1.7cm} \textcolor{teal}{\%\% V1 verifies the received SPKRSU' from RSU}\\
\scriptsize\hspace{1.7cm} \text{$/\backslash$ request(V1,RSU,auth\_2,SPKRSU')}\\
\scriptsize\hspace{1.7cm} \textcolor{teal}{\%\% V1 hopes that MSG' will be verified by RSU}\\
\scriptsize\hspace{1.7cm} \text{$/\backslash$ witness(V1,RSU,auth\_3,MSG')}\\
\scriptsize\textcolor{red}{end role}\\
\hline
\end{tabular}
\end{center}
\end{table}

\begin{table}[H]
\begin{center}
\setlength{\tabcolsep}{2pt} % Default value: 6pt
\renewcommand{\arraystretch}{0.6} % Default value: 1
\begin{tabular}{| p{8cm} |}
\hline 
\scriptsize\textbf{Code 2}\text{: HLPSL code for the role of the RSU $R_{k}$, played by $R_{k}$}\\
\hline
\scriptsize\text{\textcolor{red}{role role\_RSU} (RSU,V1:agent,PKTA,PKV1,PKRSU:public\_key,}\\
\hspace{1.45cm} \scriptsize\text{SK12:symmetric\_key,SND,RCV:channel(dy))}\\
\scriptsize\text{played\_by RSU}\\
\scriptsize\text{def=}\\
\scriptsize\hspace{0.8cm} \textcolor{blue}{local}\\
\scriptsize\hspace{1.4cm} \text{State:nat,TR,T1,T2,T3,TIDV1,SPKV1,TIDRSU,}\\
\scriptsize\hspace{1.4cm} \text{SPKRSU,MSG,PIDV1,A1:text}\\
\scriptsize\hspace{0.8cm} \textcolor{blue}{init}\\
\scriptsize\hspace{1.4cm} \text{State:=0}\\
\scriptsize\hspace{0.8cm} \textcolor{blue}{transition}\\
\scriptsize\hspace{1.4cm} \text{1. State=0 $/\backslash$ RCV(TIDV1'.SPKV1'.T1'.PKV1.TR.} \\
\scriptsize\hspace{1.4cm} \text{\{PKV1.TR\}\_inv(PKTA).\{TIDV1'.SPKV1'.T1'.PKV1.} \\
\scriptsize\hspace{1.4cm} \text{TR.\{PKV1.TR\}\_inv(PKTA)\}\_inv(PKV1)) $=|>$ State' } \\
\scriptsize\hspace{1.4cm} \text{:= 1 $/\backslash$ T2' := new() $/\backslash$ TIDRSU' := new() $/\backslash$ } \\
\scriptsize\hspace{1.4cm} \text{SPKRSU' := new() $/\backslash$ SND(TIDRSU'.SPKRSU'.T2'.} \\
\scriptsize\hspace{1.4cm} \text{PKRSU.TR.\{PKRSU.TR\}\_inv(PKTA)\}\_inv(PKRSU))} \\
\scriptsize\hspace{1.7cm} \textcolor{teal}{\%\% RSU verifies the received SPKV1' from V1}\\
\scriptsize\hspace{1.7cm} \text{$/\backslash$ request(RSU,V1,auth\_1,SPKV1')}\\
\scriptsize\hspace{1.7cm} \textcolor{teal}{\%\% RSU hopes that SPKRSU' will be verified by V1}\\
\scriptsize\hspace{1.7cm} \text{$/\backslash$ witness(RSU,V1,auth\_2,SPKRSU')}\\
\vspace{-0.03cm}\scriptsize\hspace{1.4cm} \text{2. State=1 $/\backslash$ RCV(MSG'.PIDV1'.A1'.T3'.\{MSG'.}\\
\scriptsize\hspace{1.7cm} \text{PIDV1'.A1'.T3'\}\_SK12) $=|>$ State' :=2}\\
\scriptsize\hspace{1.7cm} \textcolor{teal}{\%\% RSU verifies the received MSG' from V1}\\
\scriptsize\hspace{1.7cm} \text{$/\backslash$ request(RSU,V1,auth\_3,MSG')}\\
\scriptsize\textcolor{red}{end role}\\
\hline
\end{tabular}
\end{center}
\end{table}

\begin{table}[H]
\begin{center}
\setlength{\tabcolsep}{2pt} % Default value: 6pt
\renewcommand{\arraystretch}{0.6} % Default value: 1
\begin{tabular}{| p{8cm} |}
\hline 
\vspace{-0.1cm}\scriptsize\textbf{Code 3}\text{: HLPSL code for the roles of session and environment}\\
\hline
\vspace{-0.1cm}\scriptsize\text{\textcolor{red}{role session} (V1,RSU:agent,PKTA,PKV1,PKRSU:public\_key,} \\
\scriptsize\hspace{1.25cm} \text{SK12:symmetric\_key)}\\
\scriptsize\text{def=}\\
\scriptsize\hspace{.8cm} \textcolor{blue}{local}\\
\scriptsize\hspace{1.4cm} \text{SND1,RCV1,SND2,RCV2:channel(dy)}\\
\scriptsize\hspace{.8cm} \textcolor{blue}{composition}\\
\scriptsize\hspace{1.4cm} \textcolor{red}{role\_V1} \text{(V1,RSU,PKTA,PKV1,PKRSU,SK12,}\\
\scriptsize\hspace{2.4cm} \text{SND1,RCV1) $/\backslash$}\\
\scriptsize\hspace{1.4cm} \textcolor{red}{role\_RSU} \text{(RSU,V1,PKTA,PKV1,PKRSU,SK12,}\\
\scriptsize\hspace{2.7cm} \text{SND2,RCV2)}\\
\scriptsize\textcolor{red}{end role}\\
\vspace{-0.2cm}
\scriptsize\textcolor{red}{role environment ()}\\
\scriptsize\text{def=}\\
\scriptsize\hspace{.8cm} \textcolor{blue}{const}\\
\scriptsize\hspace{1.4cm} \text{pkta,pkv1,pkrsu:public\_key,}\\
\scriptsize\hspace{1.4cm} \text{sk12:symmetric\_key,}\\
\scriptsize\hspace{1.4cm} \text{v1,rsu:agent,}\\
\scriptsize\hspace{1.4cm} \text{auth\_1,auth\_2,auth\_3:protocol\_id}\\
\scriptsize\hspace{1.4cm} \text{intruder\_knowledge=\{v1,rsu,pkta,pkv1,pkrsu\}}\\
\scriptsize\hspace{.8cm} \textcolor{blue}{composition}\\
\scriptsize\hspace{1.4cm} \text{session(v1,rsu,pkta,pkv1,pkrsu,sk12) }\\
\scriptsize\textcolor{red}{end role}\\
\vspace{-0.2cm}
\scriptsize\textcolor{red}{goal}\\
\scriptsize\hspace{1.4cm} \text{authentication\_on auth\_1,auth\_2,auth\_3}\\
\scriptsize\textcolor{red}{end goal}\\
\scriptsize\textcolor{red}{environment()}\\
\hline
\end{tabular}
\end{center}
\end{table}

\end{document}